\documentclass[preprint,preprintnumbers,superscriptaddress,nofootinbib,showkeys]{revtex4}

\pdfoutput=1

\usepackage[warnundef]{jabbrv}

\usepackage[reqno]{amsmath}
\usepackage{amsthm}
\usepackage{amssymb}

\usepackage{graphicx}

\usepackage{multirow}

\newcommand{\etal}{{\em et~al.}}                
\newcommand{\beq}{\begin{equation}}
\newcommand{\eeq}{\end{equation}}
\newcommand{\bea}{\begin{eqnarray}}
\newcommand{\eea}{\end{eqnarray}}
\begin{document}


\title{
Symmetry Energy III: Isovector Skins
}

\author{Pawe\l~Danielewicz\email{danielewicz@nscl.msu.edu}}

\affiliation{National Superconducting Cyclotron Laboratory and
Department of Physics and Astronomy, Michigan State University,
East Lansing, Michigan 48824, USA\\
}

\affiliation{African Institute for Mathematical Sciences,
Bagamoyo, Tanzania\\}

\affiliation{Institute for Nuclear Theory, University of Washington,
Seattle, Washington 98195, USA\\
}

\author{Pardeep Singh\email{panghal005@gmail.com}}

\affiliation{National Superconducting Cyclotron Laboratory and
Department of Physics and Astronomy, Michigan State University,
East Lansing, Michigan 48824, USA\\
}

\affiliation{Department of Physics, Deenbandhu Chhotu Ram University of Science and Technology, Murthal-131039, India\\}

\author{Jenny Lee\email{jleehc@hku.hk}}

\affiliation{Department of Physics, The University of Hong Kong, Hong Kong\\}

\begin{abstract}

Isoscalar density is a sum of neutron and proton densities and isovector is a normalized difference.  Here, we report on the experimental evidence for the displacement of the isovector and isoscalar surfaces in nuclei, by $\sim$$0.9 \, \text{fm}$ from each other.  We analyze data on quasielastic (QE) charge exchange (p,n) reactions, concurrently with proton and neutron elastic scattering data for the same target nuclei, following the concepts of the isoscalar and isovector potentials combined into Lane optical potential.  The elastic data largely probe the geometry of the isoscalar potential and the (p,n) data largely probe a relation between the geometries of the isovector and isoscalar potentials.  The targets include $^{48}$Ca, $^{90}$Zr, $^{120}$Sn and $^{208}$Pb and projectile incident energy values span the range of (10-50)$\,\text{MeV}$.   In our fit to elastic and QE charge-exchange data, we allow the values of isoscalar and isovector radii, diffusivities and overall potential normalizations to float away from those in the popular Koning and Delaroche parametrization.  We find that the best-fit isovector radii are consistently larger than isoscalar and the best-fit isovector surfaces are steeper.  Upon identifying the displacement of the potential surfaces with the displacement of the surfaces for the densities in the Skyrme-Hartree-Fock calculations, and by supplementing the results with those from analysing excitation energies to isobaric analog states in the past, we arrive at the slope and value of the symmetry energy at normal density of $70<L<101\, \text{MeV}$ and $33.5<a_a^V<36.4\, \text{MeV}$, respectively.

\end{abstract}



\preprint{INT-PUB-16-036}

\maketitle


\section{Introduction}
Given the similarity between neutrons and protons, that gets ramified in the concepts of charge symmetry and charge invariance for nuclear interactions, the natural expectation is that there is little difference in how neutrons and protons distribute themselves across a~nucleus.  In fact, it has been common to infer net nuclear density by scaling up the proton density deduced from electron scattering.  Moreover, following expectations of incompressible nuclear matter, it has been common to find nuclear radii discussed as proportional to the cube root of mass number, particularly in textbooks.

Extrapolations from nuclei studied under laboratory conditions to neutron stars~\cite{horowitz_neutron_2001} spurred, though, an increased attention to potential differences in the distribution of neutrons and protons and, in particular, to the emergence of neutron skins and their interrelation with the symmetry energy, the energy penalty for neutron-proton imbalance in a nucleus.  In theoretical considerations and calculations~\cite{dobaczewski_neutron_1996,Danielewicz:2003dd}, the size of neutron skin shrinks towards zero with vanishing imbalance.  With limited relative imbalances in the larger nuclei suitable for exploring bulk nuclear properties, such as symmetry energy, the small expected sizes of the skins have made their precise determination, and the drawing of conclusions on symmetry energy, difficult.  The radii of nucleonic distributions are primarily determined in elastic scattering off nuclei, of electrons, for proton distributions \cite{de_vries_nuclear_1987}, and of protons~\cite{PhysRevC.19.1855,PhysRevC.67.054605,Shlomo19795,PhysRevC.46.1825,PhysRevC.82.044611,PhysRevC.49.2118}, alphas~\cite{PhysRevC.29.1295} and pions~\cite{Friedman:2012pa}, for neutron and proton distributions.  With a focus on parity breaking contribution to the differential cross section, the elastic electron scattering can also provide a~relatively direct information on neutron distributions~\cite{prex_collaboration_measurement_2012}, albeit at the cost of a~long running time for an experiment and elaborate efforts to suppress both statistical and systematic errors.

In our earlier work~\cite{Danielewicz200936}, hereafter referred to as I, we pointed out that the same considerations of the interaction symmetries, that produce the concept of symmetry energy, also give rise the concepts of universal isoscalar and isovector densities.  Those densities may be expected to change weakly within an isobaric chain of nuclei and may be combined to yield neutron and proton densities.  Within the Skyrme-Hartree-Fock (SHF) calculations of the density profiles for half-infinite nuclear matter in~I, we observed the isoscalar and isovector densities to vary little with the neutron-proton imbalance in the matter, particularly within the asymmetry range such as typical for heavier nuclei, consistent with the general expectations.  However, the isovector densities on its own, and in relation to isoscalar, had in these calculations some unique characteristics tied to the symmetry-energy for the employed Skyrme interactions.  In particular, when the symmetry energies exhibited a weak density-dependence at moderately subnormal densities, the isoscalar and isovector densities were similar to each other.  On the other hand, when the energies had a strong density-dependence, the isovector density tended to have its surface pushed out relative to the isoscalar density by as much as $\sim$$1 \, \text{fm}$.  Besides the slope of symmetry energy in density being correlated with the relative displacement of isovector and isoscalar surfaces, that slope was found to be correlated with a difference in steepness of the isoscalar and isovector densities, with larger slopes (or stiffer symmetry energies) yielding steeper isovector densities.  The differences between isoscalar and isovector densities could be easily understood with the Thomas-Fermi (TF) approximation \cite{Danielewicz200936}.  In fact, for a constant symmetry energy as a~function of density, the TF approximation would produce identical isoscalar and isovector densities.

In this work, we look for an evidence of the displacement of isovector surfaces relative to isoscalar, or of isovector skin, in direct-reaction cross-sections.  We further assess the extent to which other characteristics of the densities observed in I for the half-infinite matter apply to finite nuclei.  We confront our findings with results of SHF calculations for spherical nuclei and draw conclusions on the symmetry energy.  The cross-section data stem from quasielastic (QE) charge-exchange (p,n) reactions, on one hand, and from elastic proton and neutron scattering, on the other, off the same set of target nuclei: $^{48}$Ca, $^{90}$Zr, $^{120}$Sn and~$^{208}$Pb.  The~QE reactions are those to isobaric analog states (IAS) of the target ground states.  The~above processes can be jointly~\cite{patterson_energy-dependent_1976,byrd_self-consistent_1980,wong_analysis_1984,jon_analog_1997,khoa_folding_2007} described in terms of a Lane-type potential~\cite{lane_new_1962,lane_isobaric_1962} with isoscalar and isovector parts.  The components of such potential may be parametrized in a simple form \cite{patterson_energy-dependent_1976} or may be derived from folding a nucleon-nucleon (NN) potential with proton and neutron densities \cite{schery_folding_1980,khoa_folding_2007}.  In arriving at conclusions on symmetry energy, we use a relation between the profiles of the isovector and isoscalar potentials, as a proxy for the relation between the profiles of isovector and isoscalar densities.  That presumption would be generally satisfied in a folding model with the range of NN interaction idependent of the isospin channel, as commonly assumed in the literature~\cite{schery_folding_1980,khoa_folding_2007}.  In the impulse approximation, in fact, the isovector and isoscalar potentials would just represent rescaled isoscalar and isovector densities.

Regarding the past work with a scope tied to the current effort, Schery \etal\  \cite{schery_nuclear_1976,schery_folding_1980} noticed a strong sensitivity of the QE (p,n) cross sections obtained from the folding model, to the difference of the rms proton and neutron radii, currently termed neutron skin.  Using measured (p,n) cross sections they concluded on the presence~\cite{schery_p_1974,schery_nuclear_1976,schery_radius_1974} or absence~\cite{schery_evidence_1980} of the skins depending on target nuclei.  Carlson \etal\ \cite{carlson_test_1973,carlson_optical_1975} demonstrated that the description of (p,n) data can be improved when assuming a different geometry for the isovector potential, than that following from the difference between proton and neutron potentials within the popular Becchetti-Greenlees (BG) potential parametrization \cite{becchetti_nucleon-nucleus_1969}.  That was reaffirmed by Jolly \etal~\cite{jolly_$pn$_1973}.  On the other hand, Patterson \etal\ \cite{patterson_energy-dependent_1976} described a comprehensive set of (p,n) data from their measurements, in parallel with reference elastic (p,p) and (n,n) cross sections, using single geometry of the BG proton potential for both the isovector and the isoscalar potentials.  Jon \etal~\cite{jon_analog_1997,jon_isovector_2000} adjusted geometry of the imaginary part of isovector potential to reproduce their cross-section measurements for (p,n) reactions at $35 \, \text{MeV}$, while keeping the real part in the BG form, largely continuing the effort of Carlson \etal~\cite{carlson_test_1973,carlson_optical_1975}.  However, they neither considered elastic cross sections in their analysis, nor allowed for a self-consistency between the adjusted isovector potential and the initial and final wavefunctions in the employed Distorted-Wave Born Approximation (DWBA).  In our work, we find both to be of importance when deciding on the subtle details of the Lane potential, even though the contributions of the isovector potential to the nucleonic potentials are nominally small, $\sim |N-Z|/A$.
Both Jon \etal~\cite{jon_analog_1997,jon_isovector_2000} and Carlson \etal\ \cite{carlson_test_1973,carlson_optical_1975} were able to assign values of radii and surface diffuseness to isovector potentials on a nucleus-by-nucleus basis and these turned out to vary relatively smoothly with nuclear mass.  Even though we work with larger data sets with smaller errors for individual nuclei, we find those data insufficient to constrain reliably such an abundance of parameters for isovector potentials, on a meaningful scale, as in the works above.  The recent interest in the Lane potential, tied to the symmetry energy, focussed more on the strength of the isovector potential \cite{khoa_folding_2007,khoa_folding_2014,li_neutronproton_2015,liChen_neutronproton_2015} than on its geometry being central here.  In one work~\cite{loc_charge-exchange_2014}, though, ($^3$He,t) on $^{90}$Zr and $^{208}$Pb reaction data were analyzed in terms of a folding model with the goal of learning on neutron skins, with the conclusion that these met expectations.

In the next section, we discuss various practicalities in theoretical description of elastic and quasielastic processes, such as nucleon optical potential parametrizations, impact of various components on the cross sections, DWBA etc.  In Sec.~3, we fit the geometry of isovector and isoscalar potentials to data.  The QE (p,n) data we rely on are those from the measurements by Doering, Patterson and Galonsky \cite{patterson_energy-dependent_1976,doering_microscopic_1975,doering_isobaric_1974} at the incident proton energies of 25, 35 and $45 \, \text{MeV}$.  We supplement these with elastic data from the EXFOR database~\cite{otuka_towards_2014} for the same target nuclei.  Over the target nuclei, incident energies and reaction types, we exploit 48 data sets in our fits.  For exploring the potential geometry we modify the potential parametrization by Koning and Delaroche (KD) \cite{koning_local_2003}, following inspiration from~I.  In~Sec.~4, we confront the results of our data analysis with results of spherical SHF calculations for different Skyrme interactions.  Upon combining the results from that confrontation, with those from the confrontation of theoretical predictions for symmetry coefficients with data in~\cite{danielewicz_symmetry_2014}, hereafter refereed to as II, we arrive at constraints on symmetry energy and on energy of neutron matter.  For details on the symmetry energy, we refer the reader to reviews such as~\cite{li_isospin_1998}.  In arriving at the constraints we follow Bayesian inference.   We conclude in Sec.~5.

\section{Description of Elastic and Quasielastic Cross Sections}

\subsection{Optical Potential}

The Lane optical potential~\cite{lane_new_1962,lane_isobaric_1962} is of the form
\beq
\label{eq:Lane}
{ U}({\pmb r}) = { U}_0({\pmb r}) + \frac{{\pmb \tau}\,{\pmb T}}{4A} \, { U}_1({\pmb r}) \equiv { U}_0({\pmb r}) + \frac{{\tau}_3 \,{T}_3}{4A} \, { U}_1({\pmb r}) + \frac{{\pmb \tau}_\perp \,{\pmb T}_\perp}{4A} \, { U}_1({\pmb r}) \, .
\eeq
Here, $U_0$ and $U_1$ are isoscalar and isovector optical potentials, respectively, and ${\pmb \tau}$ and ${\pmb T}$ are isospin operators for the projectile nucleon and target nucleus, respectively.  The scalar product on the r.h.s.\ of \eqref{eq:Lane} can be represented as ${\pmb \tau}_\perp \,{\pmb T}_\perp = \tau_+ \, T_- + \tau_- \, T_+ $.  The factors for $U_1$ in the Lane potential are organized so that the Lane potential reduces to a familiar form when employed in the elastic channels for proton and neutron scattering, specifically
\beq
\label{eq:Unp}
U_\text{n,p} ({\pmb r}) = { U}_0({\pmb r}) \pm \frac{N-Z}{A} \, { U}_1({\pmb r}) \, .
\eeq
Here, the upper sign pertains to neutrons and lower to protons and $N$, $Z$ and $A$ refer to the target nucleus.  These potentials generally contain spin-orbit terms.  The isoscalar and isovector potentials can be conversely deduced from the nucleonic potentials with
\begin{align}
\label{eq:U0np}
U_0 ({\pmb r}) & = \frac{1}{2} \big[ U_\text{n}({\pmb r}) + U_\text{p}({\pmb r})   \big] \, , \\[.5ex]
U_1 ({\pmb r}) & =  \frac{A}{2(N-Z)} \, \big[ U_\text{n}({\pmb r}) - U_\text{p}({\pmb r})   \big] \, .
\label{eq:U1np}
\end{align}
The off-diagonal element of the Lane potential, driving the transition from the initial to final state in a (p,n) reaction, is
\beq
\langle n, Z+1| U({\pmb r}) |p, Z \rangle = 2 \frac{\sqrt{|N - Z|}}{A} \, { U}_1({\pmb r}) \, .
\label{eq:U1mtx}
\eeq

The original Lane~\cite{lane_new_1962,lane_isobaric_1962} optical potential \eqref{eq:Lane} is invariant under overall rotations in isospin space.  We will argue that Coulomb interactions, breaking the isospin invariance, produce some $Z$-dependent difference between the $U_1$-factors multiplying the third and transverse isospin components in the Lane potential.
In the general considerations of average behavior of the potential, its isoscalar part~$U_0$ can depend only on scalar quantities in isospin space so its dependence on isovector quantities is quadratic or higher, i.e.~weak.  The isovector part of the potential, that can couple to an external isospin ${\pmb \tau}$, transforms in isospin space in the same way as isospin density ${\pmb \rho}$, so can be written as that density multiplied by a scalar factor.  The~simplest scalar factor is just a constant and a constant in particular allows to meet the requirement of the potential vanishing in the absence of matter and can provide, under any circumstances, a coarse approximation to the potential changing from zero outside of matter to a finite value prevailing across the matter interior.  However, whether or not linear, the relation between the potential and density can also be weakly nonlocal.  After the values of isospin components are factored out, as components of~${\pmb T}$, the form of~$U_1$ can be retained as a profile of isospin density being multiplied by a~scalar function, to provide insights into any variations on top of the variations of isospin content.  In~II, we discussed that Coulomb interactions, while impacting the density of the third component of isospin, since displacing protons out relative to neutrons, yield no similar impact on the density of transverse isospin in an isobaric chain.  With this, one expects a difference between the $U_1$-factor multiplying the third components of isospin in \eqref{eq:Lane} and the $U_1$-factor multiplying the transverse components there. That difference should develop with growing~$Z$, mirroring the difference in the densities for isospin components.  With the isovector potential competing with isoscalar, in generating predictions for elastic differential cross sections, it can be a challenge to discern details in $U_1$ acting in elastic scattering to the level of telling them reliably from details in $U_1$ in the (p,n) reactions.  Correspondingly, for now, in the context of the data analysis, as earlier in this section we will make no distinction between $U_1$ acting in elastic reactions and that acting in quasielastic (p,n) reactions and we will rather concentrate on revealing any difference in the geometry between $U_0$ and $U_1$, using both types of reactions.  When confronting the results with structure calculations, we will get back to the consideration of different directions in isospin space, though.

In the literature, the nucleonic potentials $U_\text{p}$ and $U_\text{n}$ are commonly expressed in terms of Woods-Saxon formfactors and their derivative factors
\begin{align}
f(r,R,a) & = \frac{1}{\exp{\frac{r-R}{a}}+1} \, ,\\[.5ex]
f^d (r,R,a) & = -4 \, a \, \frac{\text{d}f}{\text{d}r} = 4\, f \, (1-f) \, .
\end{align}
For example, the popular and relatively recent parametrization of nucleonic optical potentials by Koning-Delaroche (KD) \cite{koning_local_2003} is of the form
\beq
\begin{split}
\label{eq:Usum}
U ({\pmb r}) = & -[V_V(E)+ i \, W_V(E)] \, f(r, R_V, a_V) - i \, W_D(E) \, f^d(r, R_D, a_D) \\
 & - \frac{1}{2 \, a_\text{SO} \, m_\pi^2 } \,[V_\text{SO}(E) + i \, W_\text{SO}(E)] \, \frac{1}{r} \, f^d(r,R_\text{SO},a_\text{SO}) \, {\pmb s} \, {\pmb L}
 \, .
\end{split}
\eeq
Here, ${\pmb s}$ is the spin operator for the incident nucleon and the potential strengths depend on incident energy~$E$.
For protons, the nuclear potential~$U$ is supplemented by a Coulomb potential that is conventionally taken in the form such as for a uniformly charged sphere:
\beq
V_C(r) =
\begin{cases}
\frac{Z \, e^2}{8 \pi \epsilon_0 \, R_C} \, \Big( 3 - \frac{r^2}{R_c^2} \Big) \, , & r < R_C \, ,\\
\frac{Z \, e^2}{4 \pi \epsilon_0 \, r}   \, , & r > R_C \, .
\end{cases}
\eeq

When the geometric parameter sets $(R_X, a_X)$, $X=V$, $D$, $SO$, are the same for neutrons and protons~\cite{patterson_energy-dependent_1976,varner_global_1991}, then the isoscalar and isovector potentials from~\eqref{eq:U0np} and~\eqref{eq:U1np} have the same structure~\eqref{eq:Usum} as the nucleonic potentials, with the same set of geometric parameters for the different terms in the potentials.  When the proton and neutron potentials have significant differences in their geometry, though, such as in the "best-fit" case of the BG parametrization~\cite{becchetti_nucleon-nucleus_1969}, the resulting isovector potential can have unusual structure~\cite{hoffmann_exact_1973} difficult to justify on physical grounds.  Carlson \etal~\cite{carlson_optical_1975} and Jon \etal~\cite{jon_analog_1997,jon_isovector_2000}, in fact, postulated Woods-Saxon type formfactors for the isovector potentials and adjusted their parameters directly when describing QE (p,n) cross sections.

\begin{figure}
\centerline{\includegraphics[width=.51\linewidth]{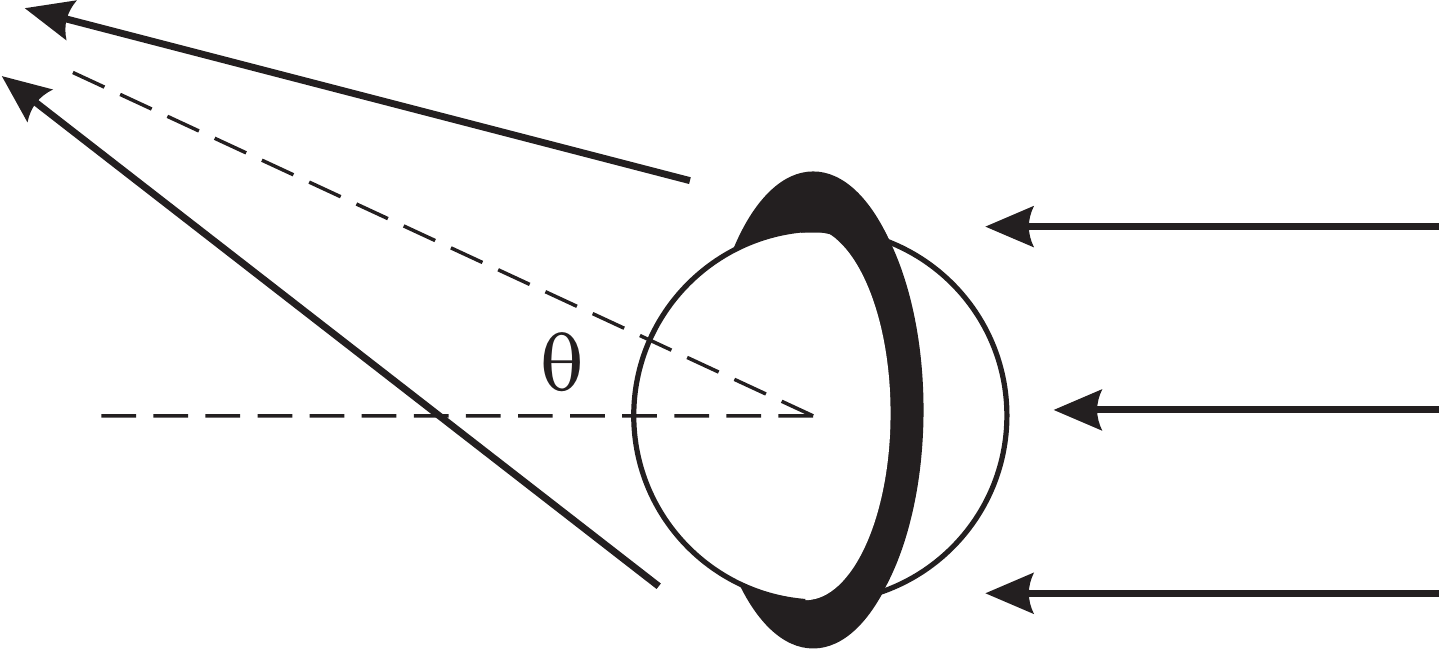}}
\caption{Schematic view of a direct reaction.
}
\label{fig:DirectReaction}
\end{figure}

As the efforts here may be of interest to those focused on symmetry energy, but whose research areas lack overlap with the direct reactions, an introduction covering essential issues of relevance for the efforts here can be found in the review by Amado~\cite{amado_analytic_1985}.  Briefly, within a direct reaction, leaving the final nucleus with little or no excitation, an important part of the nuclear processes occurs near the surface of the nucleus, with diffraction and refraction taking place there, cf.~Fig.~\ref{fig:DirectReaction}.  Traversal of the nucleus through the volume generally populates more complicated final states and depletion of the incident flux into such channels is described in terms of an imaginary and, more generally nonhermitian, part of the optical potential utilized for the direct channels, such as \eqref{eq:Usum}.
The peripheral nature of direct processes limits the complexity that can be of relevance in the potentials.

In our search for evidence of isovector skins, we use the KD potential parametrization~\cite{koning_local_2003} as the starting point.  That parametrization is relatively recent and commonly used and it is fitted to a vast collection of elastic scattering data.  Moreover it employs practically the same potential geometry for neutrons and protons.  The only small difference is in the adopted values of the diffusivity in the imaginary potential.  We actually reset that difference in the diffusivity to zero and, with the same geometry for neutrons and protons, the starting geometry becomes the same for the isocalar and isovector potentials.

\subsection{Cross Sections}

In connecting the potentials to elastic scattering data, the scattering wavefuction is decomposed in the standard manner into partial waves.  The corresponding set of one-dimensional Schr\"{o}dinger equations is solved to yield the scattering matrix elements for individual partial waves, factorized as $S_{J\ell}= \text{e}^{2 i \sigma_\ell} \, S_{J\ell}^N$, where $\sigma_\ell$ are Coulomb phase shifts and $S_{J\ell}^N$ are nuclear factors.  The cross section averaged over initial spin directions and summed up over final takes then the form
\beq
 \frac{\text{d} \sigma}{\text{d} \Omega}  =  \frac{\text{d} \sigma_C}{\text{d} \Omega} + \frac{\text{d} \sigma_N}{\text{d} \Omega} + \frac{\text{d} \sigma_i}{\text{d} \Omega} \, ,
\eeq
where $\text{d} \sigma_C/\text{d} \Omega) = |f_C(\theta)|^2$ is the Rutherford cross section, and the two remaining terms are the nuclear cross section and the interference contribution.  The interference contribution is
\beq
 \frac{\text{d} \sigma_i}{\text{d} \Omega} = 2 \, \text{Re} \left[\overline{f}_N(\theta) \, f_C^*(\theta)     \right] \, ,
\eeq
where $\overline{f}_N$ is nuclear scattering amplitude averaged over initial and summed over final spin directions
\beq
\overline{f}_N(\theta) = - \frac{i}{2k} \, \frac{1}{2s+1} \sum_{J\ell} (2J+1) \, \text{e}^{2i\sigma_\ell} \, (S_{J\ell}^N - 1) \, P_\ell(\cos{\theta}) \, ,
\eeq
where $s=\frac{1}{2}$.  Finally, the nuclear cross section is, after \cite{frobrich_theory_1996,edmonds_angular_1996,biedenharn_properties_1952},
\beq
\label{eq:sigel}
\frac{\text{d} \sigma_N}{\text{d} \Omega} = \frac{1}{k^2} \, \frac{1}{2s+1} \sum_L (2L+1) \, {\mathcal A}_L^N \, P_L(\cos{\theta}) \, ,
\eeq
where the expansion coefficients for the cross section are
\beq
\label{eq:ALel}
\begin{split}
{\mathcal A}_L^N = & \frac{1}{4} \sum_{J' \, \ell'} (2J'+1)(2\ell'+1) \sum_{J \ell} (2J+1)(2\ell+1) \, \Bigg(\begin{matrix} \ell & \ell' & L \\ 0 & 0 & 0 \end{matrix}\Bigg)^2 \\
& \times  \Bigg\lbrace\begin{matrix} \ell & \ell' & L \\ J' & J & s \end{matrix}\Bigg\rbrace^2 \, \text{Re} \left[\text{e}^{2i(\sigma_\ell - \sigma_{\ell'})}  \,  (S_{J'\,\ell'}^{N*} - 1) \,(S_{J\ell}^{N} - 1)    \right] \, ,
\end{split}
\eeq
with 3j and 6j symbols being utilized.  In elastic scattering of charged projectiles, the Coulomb cross section generally dominates in the forward direction with a~telltale sign being proximity of the net cross section to the Rutherford cross section.  The nuclear processes generally take over past the grazing angle representing the classical Coulomb trajectory that just touches the nuclear surface.  The interference contribution strongly eats into these cross sections in the vicinity of the grazing angle.  With the exception of proton elastic scattering off $^{208}$Pb at the low end of considered incident energy range, the grazing angles for the data addressed in this paper are fairly low.

The (p,n) cross sections are typically described in the literature in DWBA~\cite{frobrich_theory_1996}.  That approximation is justified when the final state is likely to be populated in one step, without other states serving as intermediaries, or multiple transitions occurring back-and-forth between the initial and final states.  Construing towards DWBA validity are fast processes, yielding low transition probability, such as typical for peripheral reactions at high incident energy.  The applicability of DWBA is expected to worsen with increasing inelasticity of a~reaction, with slowing down of the reaction, such as close to a~Coulomb barrier, and with inhibition of the probability of reaching the final state in one step, due to dynamics and/or conservation laws, as compared to a~multitude of steps, such as in emission into backward angles at high incident energies~\cite{tomita_analysis_2015}.  Multiple back-and-forth transitions between the initial and final states are usually associated with a high transition probability.  This will be addressed below.  An alternative to DWBA is a solution to a coupled channel problem that can take on various levels of complexity and physics scope.

With \eqref{eq:Lane} and \eqref{eq:U1mtx}, the unpolarized (p,n) cross section in the DWBA approximation is~\cite{frobrich_theory_1996,satchler_optical-model_1964}
\beq
\label{eq:DWBA}
\frac{\text{d}\sigma_{(p,n)}}{\text{d} \Omega} = (2 \pi)^4 \, \mu_p \, \mu_n \, \frac{k_n}{k_p} \, \frac{1}{2s+1} \sum_{M_p \, M_n} \left| 2 \frac{\sqrt{|N-Z|}  }{A} \int \text{d}{\pmb r} \, \chi^{(-)\dagger}_{n \, M_n}({\pmb r}) \, U_1({\pmb r}) \, \chi^{(+)}_{p \, M_p}({\pmb r}) \right|^2 \, ,
\eeq
where p and n are used as indices for the initial and final states, $\mu$ are reduced masses, $k$ are c.m.\ wavevectors and $\chi$ are the distorted waves that describe elastic scattering of a~proton and neutron in the initial and final channel, respectively, under the influence of the potentials \eqref{eq:Unp}.  Upon partial wave decompositions, the DWBA cross section may be brought into the form similar to that for the elastic cross section, Eqs.~\eqref{eq:sigel} and \eqref{eq:ALel}, i.e.
\beq
\frac{\text{d} \sigma_\text{(p,n)}}{\text{d} \Omega} = \frac{1}{k_p^2} \, \frac{1}{2s+1} \sum_L (2L+1) \, {\mathcal A}_L^\text{(p,n)} \, P_L(\cos{\theta}) \, .
\eeq
Here the expansion coefficients for the differential cross section are
\beq
\begin{split}
{\mathcal A}_L^\text{(p,n)} = & 4 \, \mu_p \, \mu_n \, k_p \, k_n \sum_{J' \, \ell'} (2J'+1)(2\ell'+1) \sum_{J \ell} (2J+1)(2\ell+1) \\
& \times   \Bigg(\begin{matrix} \ell & \ell' & L \\ 0 & 0 & 0 \end{matrix}\Bigg)^2 \Bigg\lbrace \, \begin{matrix} \ell & \ell' & L \\ J' & J & s \end{matrix}\Bigg\rbrace^2 \, \text{Re} \left[I_{J' \, \ell'}^* \, I_{J \ell}    \right] \, ,
\end{split}
\eeq
where $I$ are the partial-wave integrals
\beq
\label{eq:IJl}
I_{J\ell} = 2 \frac{\sqrt{|N-Z|}  }{A} \int_0^\infty \text{d}r \, r^2 \, u_{n \, J \ell}^{(+)}(r) \, U_1^{J \ell}(r) \,
u_{p \, J \ell}^{(+)}(r) \, ,
\eeq
and $u$ are radial wavefunctions for the initial and final channels.

The question whether a coupled-channel approach, combining the initial and final channels for a QE (p,n) reaction, within a Schr\"{o}dinger equation set~\cite{wesolowski_coupled-channel_1968,wong_analysis_1984,khoa_folding_2007}, could bring in some tangible benefits over DWBA, may be addressed by examining data.  In Table~\ref{tab:csratio}, we show the ratios of total QE (p,n) cross sections from the measurements of Doehring \etal~\cite{doering_isobaric_1974}, to estimated geometric nuclear geometric cross sections, $\sigma_g = \pi \, R^2 \, [1-V_C(R)/E]$.  The errors of the measured cross sections are about 10\%, but the primary uncertainty in the ratios, that may be interpreted as probabilities for p-n conversion, are tied to the ambiguity in the choice of $R$ for the geometric cross section. Here, we use the radii of the volume potential in KD parametrization, that tends to put the ratios on the high side.   As alternative to total cross section ratio, one can examine normalized differential cross sections.

\begin{table}
\caption{Percent ratio of total quasielastic (p,n) cross section from measurements of Doehring \etal~\cite{doering_isobaric_1974}, to estimated nuclear geometric cross section, $\sigma_\text{(p,n)}/\sigma_g$, where $\sigma_g = \pi \, R^2 \, [1-V_C(R)/E]$.}
\label{tab:csratio}
\vspace*{.5ex}
\begin{tabular}{||c|| c| c| c | c ||}
\hline
\hline
  & \multicolumn{4}{c ||}{Target Nucleus } \\
\cline{2-5}
 $E_p$                                      & $^{48}$Ca & $^{90}$Zr & $^{120}$Sn & $^{208}$Pb \\
 (MeV)   & ($R=4.33 \, \text{fm}$) & ($R=5.44 \, \text{fm}$) & ($R=6.03 \, \text{fm}$) & ($R=7.32 \, \text{fm}$) \\
\hline
\hline
 25 & 2.47 & 1.26 & 1.44 & 1.62 \\
 35 & 2.14 & 0.74 & 0.75 & 0.75 \\
 45 & 1.68 & 0.62 & 0.69 & 0.50 \\
\hline
\hline
\end{tabular}
\end{table}

\begin{figure}
\centerline{\includegraphics[width=.37\linewidth]{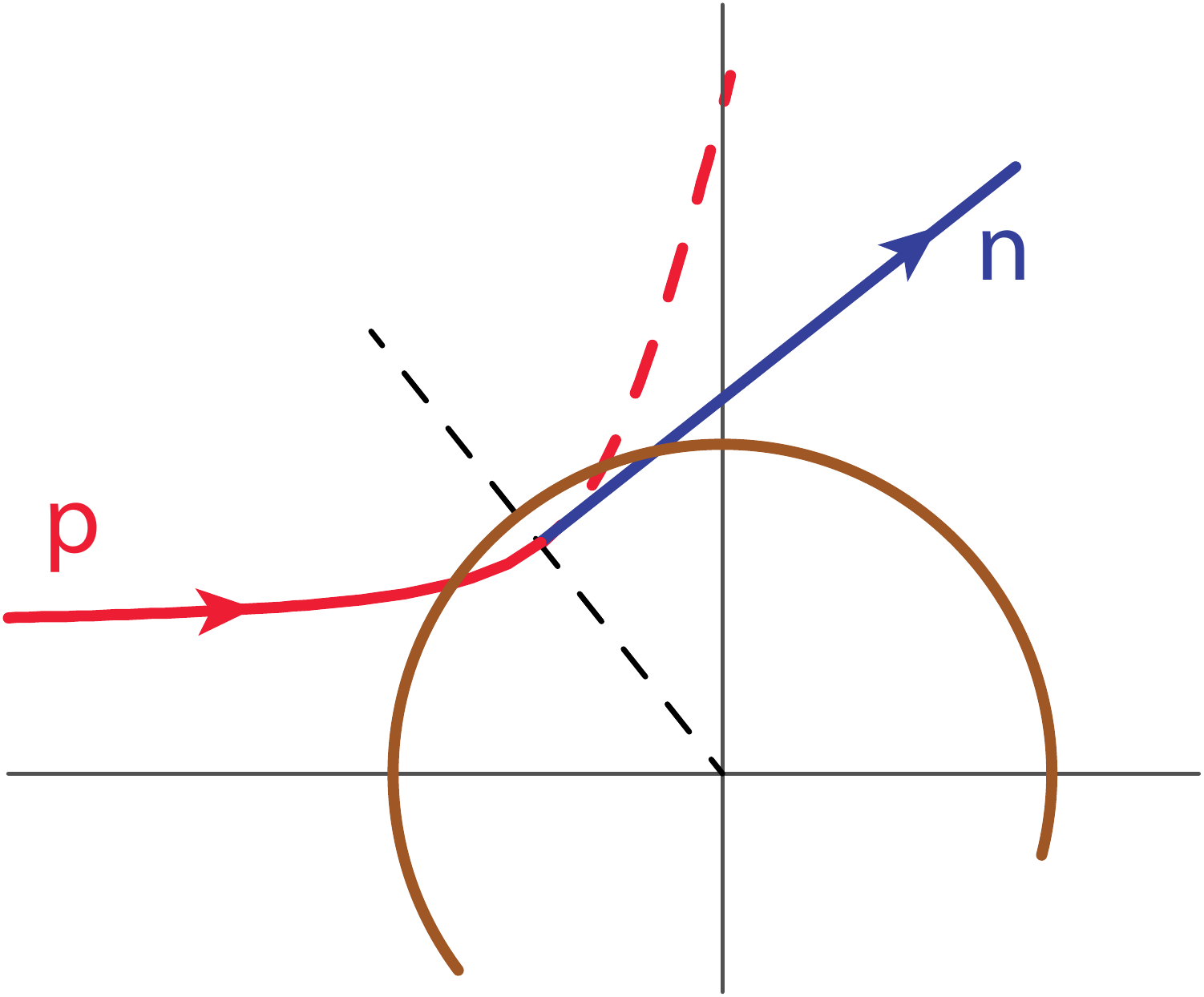}}
\caption{Charge-exchange Coulomb trajectory:  at the point of closest approach charge-exchange takes place, making the incident particle continue along an altered Coulomb trajectory towards the detectors.  The illustration here is for a (p,n) reaction on $^{208}$Pb at $25 \, \text{MeV}$ and $b= 3 \, \text{fm}$.
}
\label{fig:Coulpn}
\end{figure}

For the elastic or inelastic scattering of charged projectiles, it is common to normalize the differential cross sections with the Rutherford cross sections.  This brings in different benefits.  First, the magnitude range needed for presenting the cross section values shrinks.  Second, when diffraction effects are moderate, one can visually identify the range of angles where Coulomb scattering dominates.  Finally, when both the diffraction and refraction effects are moderate, one can interpret the ratio in terms of a transmission probability along a classical trajectory for staying within the entrance channel or for moving to another.  Normalizing the charge-exchange cross sections with the Rutherford cross-section, to reap any benefits, is pointless, though, as charges in the final state are different than in the entrance state and only half of the Coulomb deflection, behind a differential Coulomb cross section, is accumulated when the particles approach each other and the other half when they move away.  However, one might normalize a charge-exchange cross section with a Coulomb cross section arrived at under the assumption that the charge exchange occurs around the point of closest approach, with the classical trajectory illustrated in Fig.~\ref{fig:Coulpn}.  Under that assumption, the net classical deflection angle at impact parameter $b$ becomes
\beq
\theta(b) = \frac{1}{2} \, \theta_i(b) + \frac{1}{2} \, \theta_f(b) \, ,
\eeq
where the labels $i$ and $f$ pertain to the particles in the entrance and exit channels and where the r.h.s.\ angles are the Coulomb deflection angles at the impact parameter $b$ for the respective charge combinations in the entrance and exit channels.

From the above, the classical Coulomb cross section, with charge exchange at the point of closest approach, becomes
\beq
\begin{split}
\frac{\text{d} \sigma^{(i,f)}_C}{\text{d} \theta \, \, } & =  \frac{b}{\sin{\theta}} \, \frac{\text{d}  b}{\text{d}  \theta}
\\[.5ex]
& = \frac{(\kappa_i + \kappa_f)^2}{16 E^2} \, \frac{1}{\sin^4{\theta}} \, \frac{\left[\sqrt{1 - \left(\frac{\kappa_i-\kappa_f}{\kappa_i+\kappa_f}\right)^2 \, \sin^2{\theta}} + \cos{\theta} \right]^2}{\sqrt{1 - \left(\frac{\kappa_i-\kappa_f}{\kappa_i+\kappa_f}\right)^2 \, \sin^2{\theta}}} \, .
\end{split}
\eeq
Here $\kappa_{i,f}$ are Coulomb factors for the entrance and final channels, respectively, $\kappa_{i,f} = \frac{Z_{i,f} \, Z_{I,F} \, e^2}{4 \pi \epsilon_0}$.  For $\kappa_i=\kappa_f$, the above cross section becomes the standard Rutherford cross section.  For a (p,n) reaction, with $\kappa_n  = 0$, the cross section becomes
\beq
\label{eq:dsigCpn}
\frac{\text{d} \sigma^{(p,n)}_C}{\text{d} \theta \, \, }  = \frac{\kappa_p^2}{4E^2} \, \frac{\cos{\theta}}{\sin^4{\theta}} \, ,
\eeq
at $\theta < 90^\circ$ and at $\theta > 90^\circ$ the cross section is zero.  The deflection angle is half of the standard Coulomb angle for a given $b$ and the largest possible classical scattering angle, under the assumptions, is then 90$^\circ$.

\begin{figure}
\centerline{\includegraphics[width=.88\linewidth]{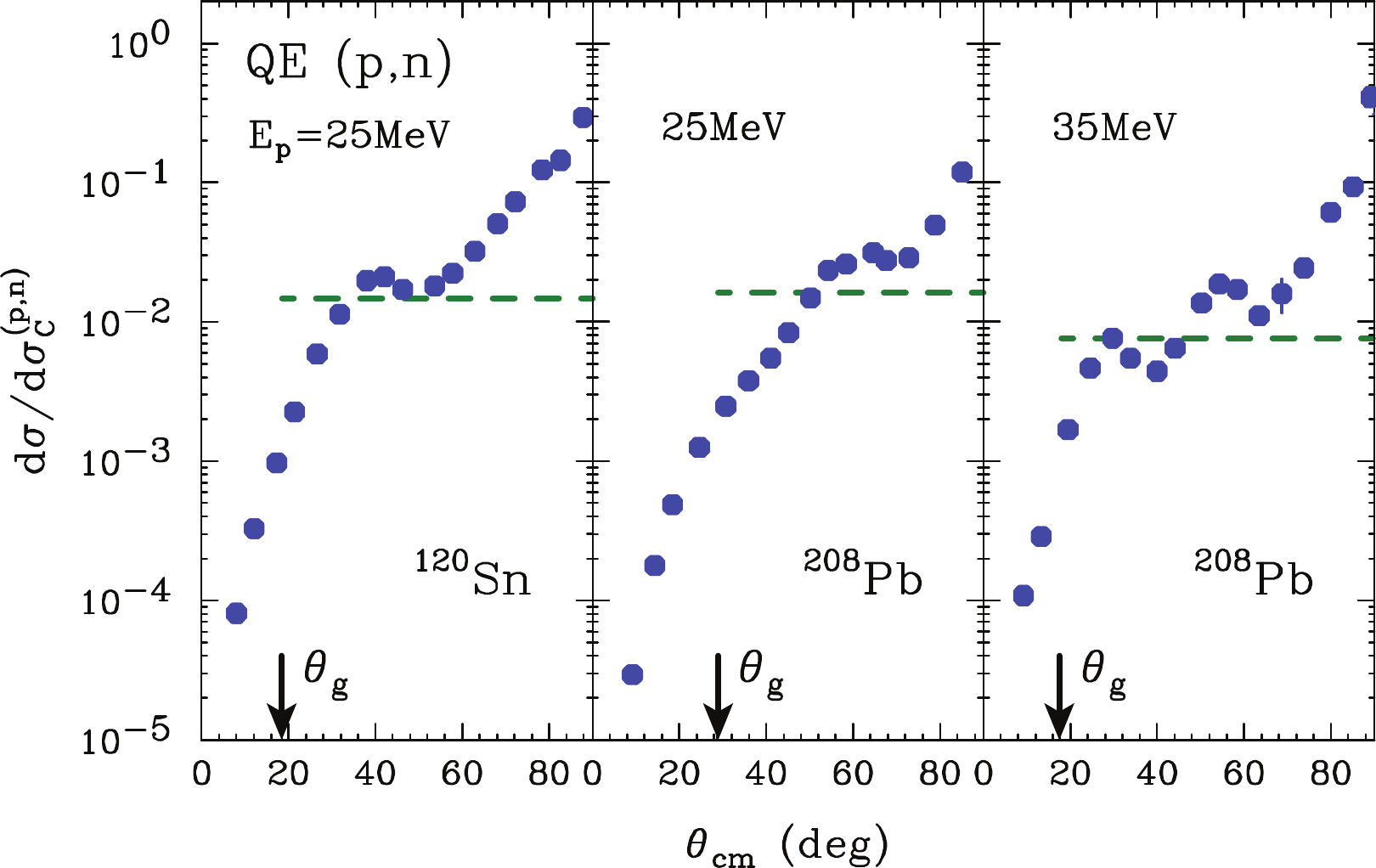}}
\caption{Differential cross sections for quasielastic charge exchange (p,n) reactions, from measurements of Doehring \etal~\cite{patterson_energy-dependent_1976,doering_microscopic_1975,doering_isobaric_1974}, normalized  with the Coulomb charge-exchange cross sections~\eqref{eq:dsigCpn}.  Dashed lines indicate total cross section ratios from Table \ref{tab:csratio}.  Arrows indicate grazing angles.
}
\label{fig:pn3R}
\end{figure}

In normalizing the measured differential cross sections with \eqref{eq:dsigCpn}, we expect low values of the cross section ratio at angles lower than the grazing angle $\theta_g$ (half of the grazing angle for Rutherford scattering).  Emission into those angles is actually most likely to origin from lower angular momenta than those expected for classical Coulomb trajectories populating the angles.  Towards 90$^\circ$, the ratio is going to diverge, as refraction and diffraction due to nuclear processes have no problem populating the wide angles, towards 90$^\circ$ and beyond.  Just past the grazing angle we expect cross-section ratios from the data that can genuinely reflect probabilities from moving from the entrance on local classical trajectories.  In Fig.~\ref{fig:pn3R}, we show the (p,n) cross sections from the measurements of Doehring \etal~\cite{patterson_energy-dependent_1976,doering_microscopic_1975,doering_isobaric_1974}, normalized with Coulomb (p,n) cross section, for systems that combine large target mass with low incident energy.  For those systems the spread of classical trajectories into wide angles is greater and, in the ratios, one can observe plateaus past the grazing angles.

The differential ratios in the plateau regions in Fig.~\ref{fig:pn3R} remain of the same order of magnitude as those in Table~\ref{tab:csratio}.  We indicate the ratios from the table with dashed lines in Fig.~\ref{fig:pn3R}.  Notably the integral of the Coulomb cross section, from the grazing angle on, is equal to the geometric cross section, $\int_{\theta_g} \text{d}\sigma_C^\text{(p,n)}=\sigma_g$.  Small values of the transition probability between the channels indicate that any transition iterations between channels are not likely and the single step, represented by DWBA, should dominate.  There is one caveat to the values in Table~\ref{tab:csratio} or in Fig.~\ref{fig:pn3R}, though, in that they actually reflect a product of (p,n) conversion probability and transition probabilities for getting in and out of the region where the conversion can take place.  The latter can be in particular assessed from the mentioned ratios of elastic proton cross sections to Coulomb cross sections and are relatively large in the forward direction, of the order of one over few.  With this, the corrected estimated local transition probabilities are still fairly low, of the order few percent.

\subsection{Influence of Potentials on Cross Sections}

\begin{figure}
\centerline{\includegraphics[width=.65\linewidth]{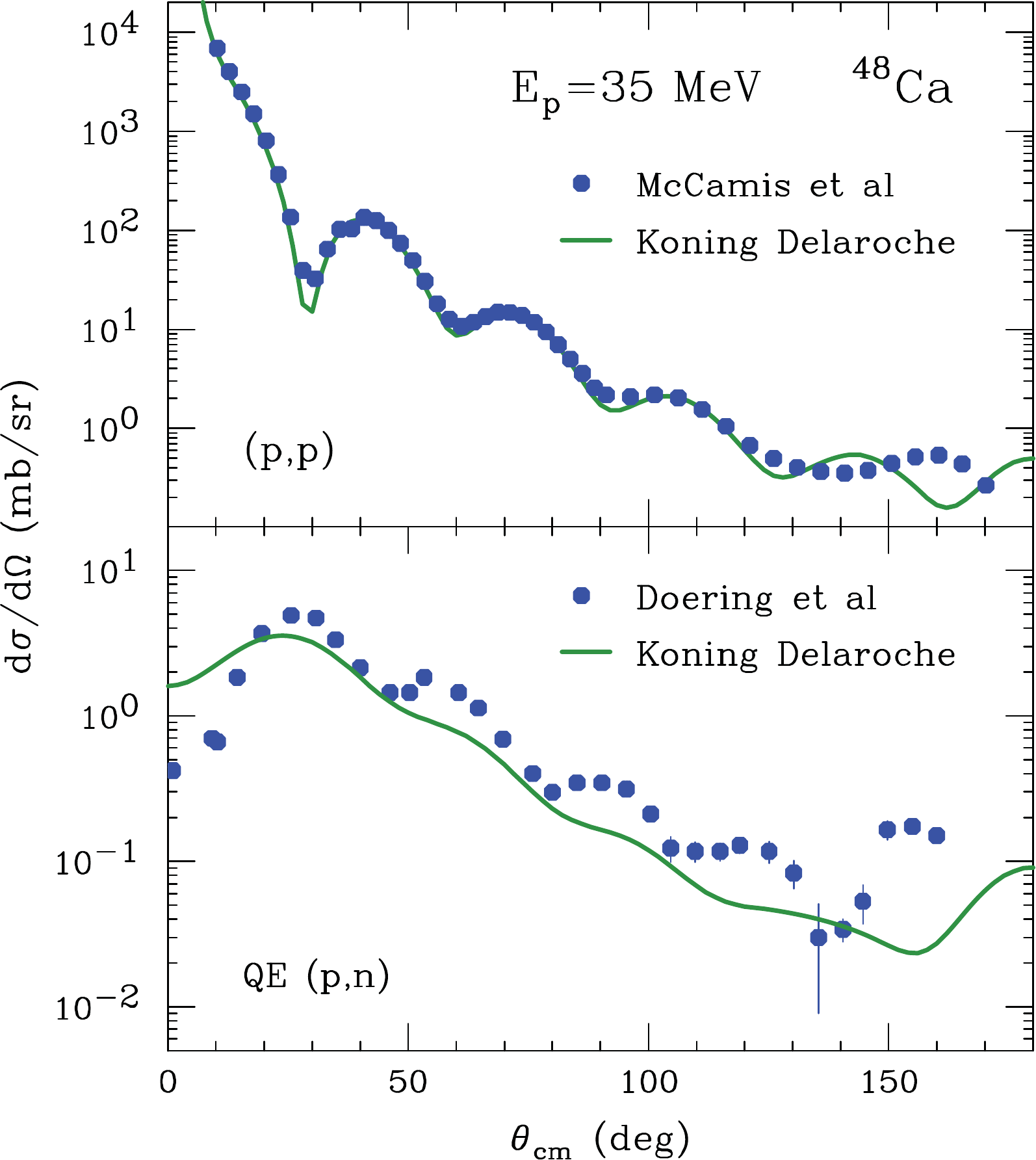}}
\caption{Differential cross section in elastic proton scattering (top panel) and quasielastic (p,n) charge-exchange reaction (bottom) on $^{48}$Ca at the incident energy of $35 \, \text{MeV}$.  Filled circles represent measurements of McCamis \etal~\cite{mccamis_elastic_1986} and Doering \etal~\cite{doering_microscopic_1975}, respectively, for the two reactions.  Solid lines represent calculations based on the KD potential \cite{koning_local_2003}.
}
\label{fig:ppn48Ca35}
\end{figure}

For light projectiles at low incident energies, the differential cross sections for direct processes are significantly impacted by wave phenomena, evident, in particular in strong oscillations of these cross sections with emission angle, see the examples in Fig.~\ref{fig:ppn48Ca35}.  The~major factor behind these oscillations is  the interference of contributions to the scattering amplitude from different regions of the target surface, see Fig.~\ref{fig:DirectReaction}, in particular the near side and far side for any given scattering angle $\theta$.  Coarsely, the positions of maximae and minimae in the interference pattern are tied to the mean radius of the bright ring on the surface shining towards the detectors. (We may notice in Fig.~\ref{fig:ppn48Ca35} a similar periodicity in the angle for the differential elastic and QE cross sections.)  The volume of the nucleus contributes to the amplitude and the cross section as well, but it yields a more smudged out pattern.  In the angular region where the wave phenomena dominate the cross section, i.e.\ wide angles, the general fall-off of the cross section with angle is tied to the sharpness of edges of the emitting regions~\cite{amado_analytic_1985}, with a sharper edge leading to a slower fall-off and softer to faster.

\begin{figure}
\centerline{\includegraphics[width=.58\linewidth]{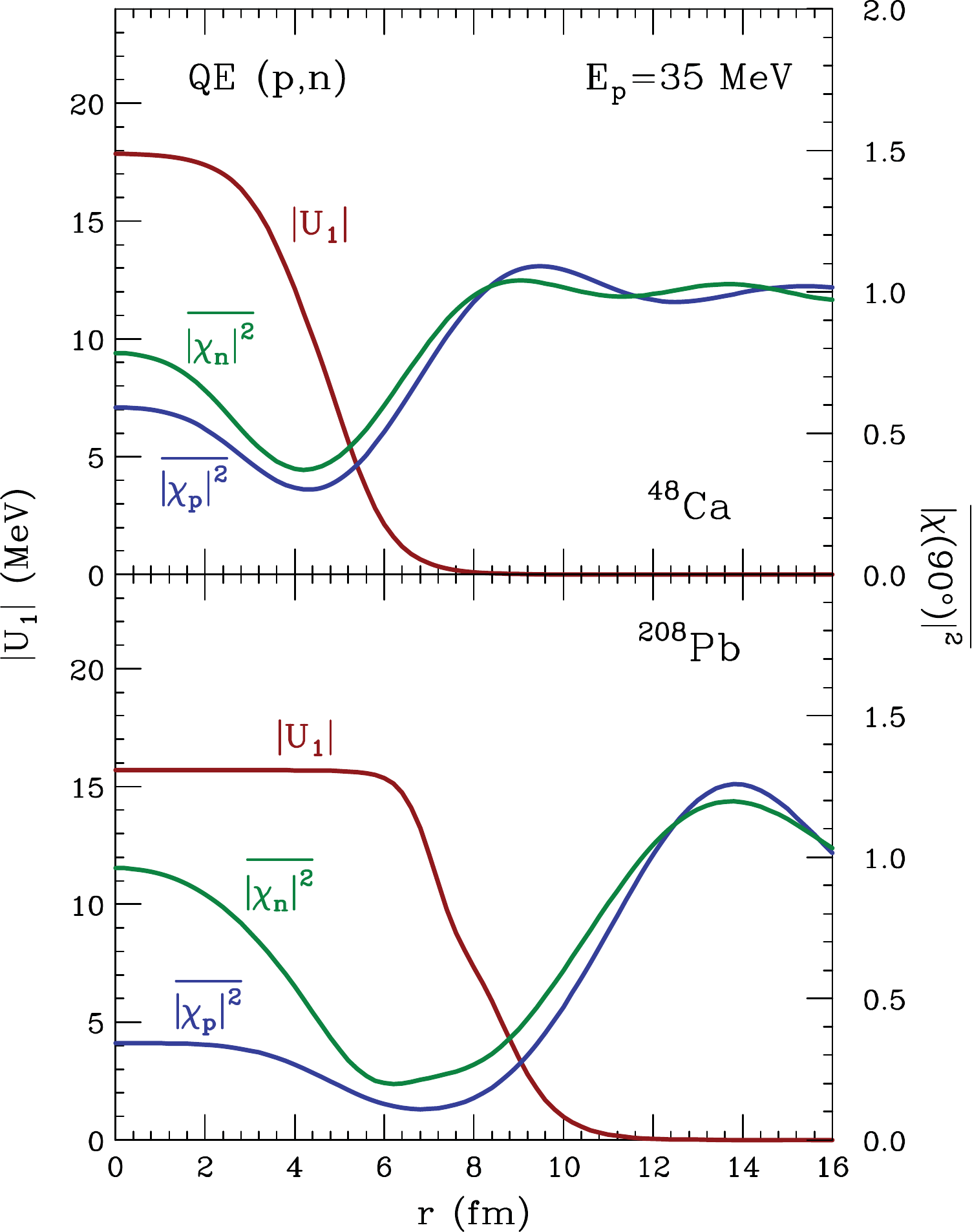}}
\caption{Spin-averaged square moduli of the distorted wavefunctions describing initial and final states in the quasielastic (p,n) reactions at the incident energy of $35 \, \text{MeV}$, on selected nuclei, as well as the modulus of isovector potential, as a function of the distance $r$ from target center.  The results here are arrived at when utilizing the KD parametrization for the optical potential.  Direction of the position vector here, for the respective wavefunctions, is taken as perpendicular to the initial and final wavevectors.
}
\label{fig:wfiu35}
\end{figure}

Figure~\ref{fig:wfiu35} further provides illustration for
the specific situation of the emission region in the context of the DWBA amplitude for a (p,n) reaction in Eq.~\eqref{eq:DWBA} with the KD parameterization of optical potential.  Besides the magnitude of the transition potential $|U_1|$, to be further discussed, we show there the spin-averaged moduli of distorted wavefunctions, that can be obtained following expressions similar to Eqs.~(\ref{eq:DWBA})-(\ref{eq:IJl}):
\beq
\overline{|\chi({\pmb r})|^2} = \frac{1}{2s+1} \sum_{M_s} \chi_{M_s}^\dagger({\pmb r}) \, \chi_{M_s}({\pmb r}) = \frac{1}{2s+1} \sum_L (2L+1) \, {\mathcal A}_L^\rho(r) \, P_L(\cos{\theta}) \, .
\eeq
Here $\theta$ is the angle between wavevector ${\pmb k}$ and position vector ${\pmb r}$ for the wavefunction and
\beq
\begin{split}
{\mathcal A}_L^\rho = &  \sum_{J' \, \ell'} (2J'+1)(2\ell'+1) \sum_{J \ell} (2J+1)(2\ell+1) \\
& \times   \Bigg(\begin{matrix} \ell & \ell' & L \\ 0 & 0 & 0 \end{matrix}\Bigg)^2 \Bigg\lbrace \, \begin{matrix} \ell & \ell' & L \\ J' & J & s \end{matrix}\Bigg\rbrace^2 \, \text{Re} \left[i^{\ell - \ell'}  \,  u_{J' \, \ell'}^*(r) \, u_{J \ell}(r)    \right] \, .
\end{split}
\eeq
For $r \rightarrow \infty$, the wavefunction moduli squared $|\chi|^2$ tend to~1.  The final-state wavefunction for the DWBA approximation is computed using the $U_n$ potential from the KD-parametrization, at the equivalent incident energy for the final-state neutron.

Regarding the isovector potential displayed in Fig.~\ref{fig:wfiu35}, the energy dependence in a potential, such as within the potential strengths the KD parametrization, complicates the situation, compared to that anticipated for Eqs.~\eqref{eq:U0np} and~\eqref{eq:U1np} for arriving at the isoscalar and isovector potentials.  This is because a proton projectile slows down due to Coulomb repulsion when approaching the target nucleus~\cite{patterson_energy-dependent_1976,devito_neutron_2012}, making it inappropriate to combine the p and n potentials for the same asymptotic energy.  This issue is reflected in the Q-values for QE (p,n) reactions being dominated by the Coulomb displacement energies and thus reduced energies of the emerging neutrons compared to the incident protons.  In the literature, e.g.~\cite{patterson_energy-dependent_1976,devito_neutron_2012}, this has been ameliorated by reducing the neutron energy, compared to proton, for the nucleonic potential in the equations such as~\eqref{eq:U0np} and~\eqref{eq:U1np}.  We account for the slowdown by determining the isoscalar and isovector potential components from the combination of entrance proton and exit neutron potentials in the (p,n) reaction, rather than the potentials on a single nucleus at one incident energy for neutron and proton.
With p and n referring then both to the particle and the channel in a (p,n) reaction, we determine then $U_0$ and $U_1$, from $U_p$ and $U_n$, with
\begin{align}
\label{eq:UE1np}
U_1 ({\pmb r}) & =  \frac{A}{2(N-Z-1)} \, \big[ U_\text{n}({\pmb r}) - U_\text{p}({\pmb r})   \big] \, , \\[.5ex]
U_0 ({\pmb r}) & =  U_\text{p}({\pmb r}) + \frac{N-Z}{A} \, U_1({\pmb r}) \,  ,
\label{eq:UE0np}
\end{align}
rather than from \eqref{eq:U0np} and~\eqref{eq:U1np}.  Here, $Z$ and $N$ refer to the target nucleus in the reaction and the incident energies in the nucleonic potential parametrizations represent the two channels.

  The bulk of the nuclear potentials that shape, cf.~Eq.~\eqref{eq:Unp}, the wavefunctions  illustrated  in Fig.~\ref{fig:wfiu35}, is isoscalar.
In that figure one can see a depletion of the wavefunctions in the target region, for proton and neutron channels, stronger for $^{208}$Pb than $^{48}$Ca.  The depletion is primarily due to the loss of probability flux to other channels and described by the imaginary part of the optical potential.  Some of the depletion is due to acceleration, compared to the asymptotic region, as nucleons enter attractive nuclear potential.  Finally, for protons some of the depletion is due to the deflection by the outside Coulomb potentials.  Notably the reduction in the probability density, compared to the asymptotic regions is by a factor of few at most, consistent with the claim before, that the transmission probability into the inner nuclear region is of the order of one over few.
The enhancement in the probability densities at the center of nuclei is due to the waves coming from all sides from the surface and interfering constructively around $r = 0$.  That enhancement is analogous to that in Arago-Fresnel-Poisson spot at the center of a circular shadow.  The spatial extent of the enhancement shrinks as the projectile energy increases.  Intuitively, at one incident energy, overall central enhancement is expected to be weaker for a  target larger in size.  The situation is reversed here at one incident energy for the two reactions due to the stronger slowing down of the projectile proton by the Coulomb potential for the larger target.

The surface ring contributions to the charge-exchange amplitude \eqref{eq:DWBA}, illustrated in Fig.~\ref{fig:DirectReaction}, stem from an interplay of the fall-off of nucleonic wavefunctions entering the nucleus and the isovector potential $U_1$ declining with the exit from a nucleus.  In Fig.~\ref{fig:wfiu35}, it is apparent that the product of $U_1$ and the proton and neutron wavefunctions is not going to produce a significant surface enhancement in the production amplitude for the original KD parametrization of the optical potential.  In the calculated (p,n) cross section in Fig.~\ref{fig:ppn48Ca35}, it is indeed observed that interference oscillations are very weak in disagreement with the data~\cite{doering_microscopic_1975}.  Contrasting further with the situation for the charge-exchange reaction, the KD optical-potential parametrization yields a very good description of elastic data at the same incident energy of $35\, \text{MeV}$ as the discussed charge-exchange, cf.~in particular Fig.~\ref{fig:ppn48Ca35}.

Figure~\ref{fig:wfiu35} suggests, though, that the situation of describing charge-exchange data could be much improved, at least qualitatively, if the isovector potential were characterized by a~bit larger radii than the isoscalar.  A resulting enhancement in the surface contribution to the amplitude~\eqref{eq:DWBA} could then plausibly produce interference oscillations comparable to the data.  At the same time, given that the isovector potential enters the proton potential \eqref{eq:Unp} suppressed by the asymmetry factor, the quality of the description of elastic data might not change much.

\begin{figure}
\centerline{\includegraphics[width=.65\linewidth]{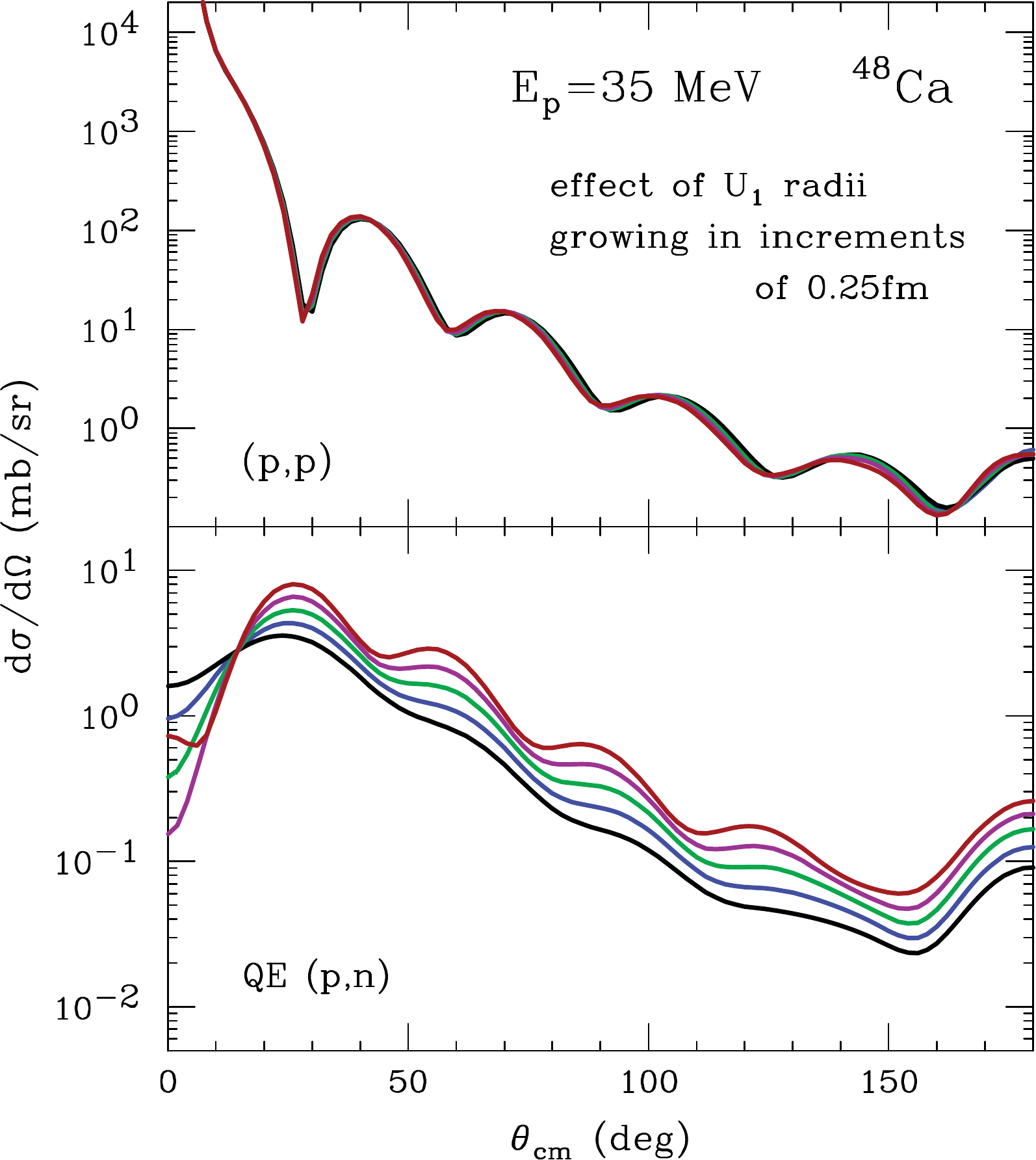}}
\caption{Differential cross sections in elastic proton scattering (top panel) and quasielastic (p,n) charge-exchange reaction (bottom) on $^{48}$Ca at the incident energy of $35 \, \text{MeV}$, calculated while increasing radii in the isovector component $U_1$ of the KD optical potential parametrization in $0.25 \, \text{fm}$ increments, from 0 to $1 \, \text{fm}$.  For both cross sections the pattern of oscillations generally shifts towards smaller angles and the magnitude of oscillations increases as the radii increase.
}
\label{fig:ppn48Ca35_fvr}
\end{figure}

The above idea is next tested in Fig.~\ref{fig:ppn48Ca35_fvr}.  We show there elastic and QE cross sections generated when using a modified KD parametrization, with enhanced isovector radii.  In the modification, the isoscalar potential is kept intact (Eq.~\eqref{eq:UE0np}), while the isovector potential is generated assuming that the radii in the volume, surface and spin-orbit terms of the nucleonic potentials are all enhanced by $\Delta R_1$ when used in~Eq.~\eqref{eq:UE1np}.  The enhancement cycles through the values $\Delta R_1 = 0, 0.25, 0.5, 0.75$ and $1 \, \text{fm}$,.  The potential depths and diffuseness are kept intact.  The nucleonic potentials, for arriving at the elastic cross section and the initial and final states in DWBA, are then constructed from the isoscalar and the modified isovector potentials by combining those potentials in the standard manner in Eq.~\eqref{eq:Unp}.  For $\Delta R_1 = 0$, the procedure just restores the original nucleonic potentials.  For an identical starting Woods-Saxon geometry in the two nucleonic potentials, the procedure increases the isovector potential radii by $\Delta R_1$ and keeps the isoscalar radii identical to the original nucleonic radii.

It may be surprising that we test out values of $\Delta R_1$ in Fig.~\ref{fig:ppn48Ca35_fvr}, that are relatively large compared to the typical expectations regarding neutron skins in nuclei.  For one, the neutron skins are typically quantified as a difference in the rms radii for neutrons and protons, and we address here a displacement of the surface radii, larger than the rms radii by a factor of $\sqrt{5/3}$ in the uniform-sphere model.  Second, the difference in rms radii for neutrons and protons can be expressed in terms of the difference in rms radii for isovector and isocalar densities and, in that relation, the neutron skins are reduced by asymmetry factor:
\beq
\begin{split}
\langle r^2 \rangle_n^{1/2} - \langle r^2 \rangle_p^{1/2} & = \frac{N^2 - Z^2}{2 N Z} \, \Big(  \langle r^2 \rangle_1^{1/2} - \langle r^2 \rangle_0^{1/2}  \Big) \, \bigg[ 1 + {\mathcal O} \bigg( \frac{\langle r^2 \rangle_1^{1/2} - \langle r^2 \rangle_0^{1/2}}{\langle r^2 \rangle_0^{1/2}}    \bigg)       \bigg] \\
& \simeq 2 \, \frac{N-Z}{A} \, \Big(  \langle r^2 \rangle_1^{1/2} - \langle r^2 \rangle_0^{1/2}  \Big) \, .
\end{split}
\eeq
Here, the subscripts 0 and 1 refer to the rms radii computed with isoscalar and isovector densities, respectively, and the last result is obtained assuming that both the difference in radii and the asymmetry are small.  For $^{48}$Ca, the combined amplification factor, of the surface isovector skin over the neutron rms skin, is of the order of 4 and, for $^{208}$Pb, it is of the order of 3.  In the latter case, an additional dynamic source of difference between the skins are significant Coulomb forces that differently impact different directions in isospin space.

It is apparent in Fig.~\ref{fig:ppn48Ca35_fvr} that changes in $U_1$ radii have a strong impact on the charge-exchange cross section.  Consistently with the expectation above, oscillations in the cross section become more pronounced with an increase in $\Delta R_1$.  Separations in the angle between enhancements in the cross section decrease with the increase, consistently with a growth in the average radius of the surface ring emitting neutrons.  If we simultaneously increase radii for $U_1$ and decrease for $U_0$,  by the same amount, the separations between the enhancements stay about the same as the average radius of the ring stays about the same.  On the other hand, the effect of changing $U_1$ radii on elastic proton cross section in Fig.~\ref{fig:ppn48Ca35_fvr} is rather minute. This is particularly shocking in that the proton elastic scattering cross sections have been used to tell the size of the neutron skin~\cite{Shlomo19795,PhysRevC.82.044611,karataglidis_discerning_2002}.  Even with the effect being slight, as $\Delta R_1$ increases, the separation between maximae in proton cross section decreases. This can be attributed to the fact that for protons the $U_1$ contributions to the real and imaginary potential come in with the same sign as the dominating $U_0$ contributions and, thus, effectively an increase in $\Delta R_1$ pushes out  the overall nuclear potential range.

\begin{figure}
\centerline{\includegraphics[width=.65\linewidth]{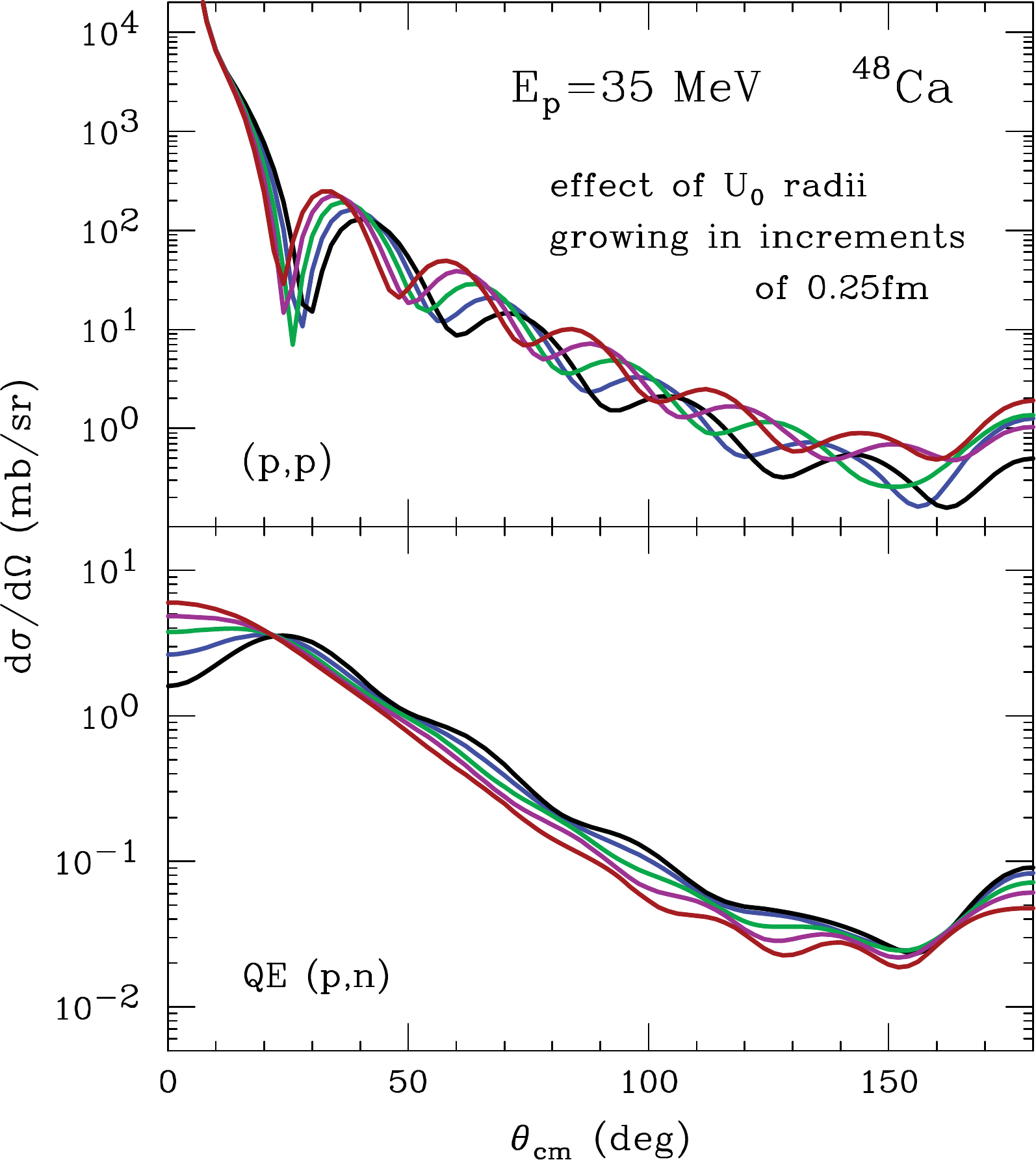}}
\caption{Differential cross sections in elastic proton scattering (top panel) and quasielastic (p,n) charge-exchange reaction (bottom) on $^{48}$Ca at the incident energy of $35 \, \text{MeV}$, calculated while increasing radii in the isoscalar component $U_0$ of the KD optical potential parametrization in $0.25 \, \text{fm}$ increments, from 0 to $1 \, \text{fm}$.  For the elastic cross section the pattern of oscillations generally shifts towards smaller angles and the magnitude of the oscillations increases as the radii increase.  For the quasielastic cross section the opposite is true.
}
\label{fig:ppn48Ca35_fsr}
\end{figure}

We next show in Fig.~\ref{fig:ppn48Ca35_fsr} the impact on the cross sections of increasing the isoscalar potential radii.  In analogy with the previous case, the isoscalar potential is constructed using the KD parametrization of nucleonic potentials modified by increasing the radii in the potentials on the r.h.s.\ of Eq.~\eqref{eq:UE0np} by the values of $\Delta R_0 = 0, 0.25, 0.5, 0.75$ and $1 \, \text{fm}$.  The~isovector potential, on the other hand, is taken from the KD parametrization without any modification.  The nucleonic potentials are then constructed in the standard manner from Eq.~\eqref{eq:Unp}.

We can see in Fig.~\ref{fig:ppn48Ca35_fsr} that the changes in the $U_0$ radii have a rather dramatic effect on the elastic scattering cross section.  As the radii grow, the maximae in the scattering cross section move forward in the angle and the separations between the maximae decrease.  The~changes in the $U_0$ radii also impact QE cross sections.  Those changes, though, are opposite to the changes generated by the rise in the $U_1$ radii.  Specifically, as the $U_0$ radii grow the oscillations in the charge-exchange cross section in the forward hemisphere dim out.  This is consistent with the inner radius of the shining surface ring, cf.~Fig.~\ref{fig:DirectReaction}, growing and thus the ring thinning to extinction.  It is obvious that the overall strength of the oscillations in a~charge-exchange cross section, particularly relative to the whole cross section, will depend on the difference of displacements in radii, $\Delta R = \Delta R_1 - \Delta R_0$, when the both sets of radii, isovector and isoscalar, get varied.  Separations between the enhancements in the cross section at wide angles will coarsely depend, on the other hand, on the average of the two displacements.

\section{Data Interpretation}

\subsection{General Strategy}
\label{ssec:strategy}

Following the preliminary assessment, we turn to a systematic examination of elastic and quasielatic cross sections, with the goal of quantifying possible geometric differences between isovector and isoscalar potentials and, by proxy, isovector and isoscalar densities.  The~data that we attempt to describe stem from elastic proton and neutron scattering and (p,n) QE reactions on the following nuclei: $^{48}$Ca, $^{90}$Zr, $^{120}$Sn and $^{208}$Pb.   The (p,p) data stem from the similar incident energy range as the (p,n) reactions and the (n,n) data stem from the incident energy range similar to that for the neutrons in the final states of the analyzed (p,n) reactions.

In describing the data, our starting potential is that from the KD parametrization, minimally modified to make the proton and neutron potential geometry exactly the same and, thus, yielding exactly the same starting geometries for the isoscalar and isovector potentials.  In fitting the data, we depart from those starting geometries to exhibit a preference of the data, or lack thereof, for the differences in the two geometries, inspired by the theory in~I.  Rationale for specifics of our strategy is discussed below.

From the two types of cross sections, the QE cross sections and, in particular, the oscillations in that cross section best reflect the relative displacement of the isovector and isoscalar surfaces.  Both the elastic and QE cross sections reflect geometric characteristics of the isoscalar potential.  However, measurements of protons tend to be far more accurate than of neutrons, while spanning a larger range of energies and angles, so data on proton elastic scattering can be more effective in constraining the isoscalar potential than can the data on either QE charge-exchange reactions or on neutron elastic scattering.  Along that line, one might insist on determining the isovector potential from proton elastic cross sections as well, but, when the impact of a potential on the cross sections is weak, one may worry about stability of the inversion procedure leading from the data to the deduced potential parameters.  Even when the statistical measurement errors are small, there will be systematic errors present  and these may be amplified by proximity to an instability in the inversion, there when a correlation between the parameters and data is weak.  The systematic errors include those on the theory side, such as the use of the local form of isoscalar potential and the presumed details in the shapes of potential components.

The accuracy in neutron measurements interplays with the energy of the outgoing neutrons.  When data is limited in terms of the energy span for the reactions, selection of outgoing angles and measurement accuracy, there is a limit on the number of parameters that may be reliably extracted from the data.  In deciding on what information to extract, we combine the past experiences in the field, with a guidance from the theory.  Thus, it has been possible in the past to describe elastic scattering data using a geometry of potential components that did not change with the incident energy.  We assume this to hold as well for the isovector as for the isoscalar potential components.  The theory suggests~(cf.~I) that isovector density is pushed out from a nucleus relative to isoscalar density, and this is of interest here, so we need to allow for the relative displacement of isovector and isoscalar radii.  However, that relative displacement may depend on the specific potential component.  E.g.~within a Thomas-Fermi consideration, the isovector and isoscalar densities and the dependence of symmetry energy on isoscalar density all interplay in generating the real part of the isovector potential.  However, the outer parts of both the isoscalar and isovector potentials are primarily imaginary and impact oscillations in the cross sections the most.  Even if their connection to the actual respective density is nonlocal, there is no reason to believe that the range of the nonlocality is significantly different for the two potentials.  To refrain from making fits unstable, we assume the same displacement for all potential components and accept that the conclusions on the displacement will primarily pertain to the absorptive part of the potential.  Still we allow the displacements to be different for individual isobars in the reactions we analyze.  There might be diferent shell effects involved when moving from one isobar to another and, further, the universality of the displacements or lack thereof may tell whether we could be accessing a~genuine pronounced physical effect or whether we might be just adding a~fit parameter with no clear support from the data.  Since the radii for nucleonic potentials in the original KD parametrization combine the effects of isovector and isoscalar potentials, we allow for a finite displacement of the isoscalar radii from the original KD parameterization, again the same for all potential components for a given isobar.  Again, there might be shell effects playing a~role and, second, for getting a good description of proton elastic cross sections, any change in isovector radii may require an opposite, but smaller in magnitude, change in the isoscalar radii.  Moreover, we allow for changes in the diffusivities for both isoscalar and isovector potentials, compared to starting values.  However, we found that the neutron data have a~too weak constraining power to place reliable constraints on the adjustments in diffusivity of isovector potential on an isobar-by-isobar basis.  In~consequence, we allow for global adjustments in the diffusivity across different components of isoscalar potentials for the individual isobars, but allow only for one universal shift in diffusivity across different components of isovector potential, relative to the isoscalar components, across all the analyzed isobars.

Besides geometry, the cross sections are obviously also sensitive to potential strengths.  In~fact, coarsely the magnitudes of cross sections are often said to reflect the volume integrals of optical potentials.  If we change radii, we need to allow for changes in potential magnitudes to permit a sensible description of the data.  Since we allow only for a single universal adjustment of the radius per potential type across an isobar, isoscalar and isovector, we consistently allow only for an adjustment in the strength of that potential type by a single factor across that isobar.

In the end, the adjusted parameters per isobar for isoscalar potentials are the displacements of radii $\Delta R_0$ and of diffusivities $\Delta a_0$ and the strength renormalization ${\mathcal F}_0$, all relative to the KD parametrization where the diffusivity of the surface imaginary potential was set to a representative value of $a_D = 0.570 \, \text{fm}$, rather than varied by up to $\sim 0.1 \, \text{fm}$ with target mass depending on projectile~\cite{koning_local_2003}.  The per-isobar-adjusted parameters for isovector potentials are the displacements of radii {\em relative to isoscalar potentials}, $\Delta R = \Delta R_1 - \Delta R_0$, and modifications of strengths {\em relative to modifications of isoscalar potentials} ${\mathcal F} = {\mathcal F}_1/{\mathcal F}_0$.  I.e.~the displacements of isovector radii relative to the original KD parametrization become equal to $\Delta R_1 = \Delta R_0 + \Delta R$, and strength modifications become ${\mathcal F}_1 = {\mathcal F} \, {\mathcal F}_0$.  Finally, we adjust a~global, across target masses, displacement of diffusivity for isovector potentials {\em relative to isoscalar potentials} $\Delta a = \Delta a_1 - \Delta a_0$.  The changes in diffusivities relative to the starting parametrization are, similarly to the case of radii, $\Delta a_1 = \Delta a_0 + \Delta a$.  In some fits, to be discussed, we required additional parameters, such as $\Delta R$, to be mass independent.

\subsection{Data Choice}

In drawing conclusions on relative geometry for isovector and isocalar potentials we exploit data on QE (p,n) reactions on $^{48}$Ca, $^{90}$Zr, $^{120}$Sn and $^{208}$Pb, at incident proton energies of 25, 35 and $45 \, \text{MeV}$, from the measurements by Doering \etal.~\cite{patterson_energy-dependent_1976,doering_microscopic_1975,doering_isobaric_1974}.  Their advantage is the span over angles, incident energies and target nuclei, all arrived at within the same experimental setup, eliminating for us the issue of a relative normalization when different data are combined in analysis.
We tried to exploit results of other measurements of the (p,n) reactions on the same target nuclei, contained the EXFOR database \cite{otuka_towards_2014}, but they did not seem to seriously augment the information beyond that from the measurements by Doering \etal , due to large errors and/or sparse angular coverage.

We complement the experimental (p,n) results with those from elastic proton and neutron scattering off the same target nuclei as in the (p,n) reactions, requiring that the particle incident energy matches the general energy region for, respectively, either p or n in the quasieleastic (p,n) reaction.  To find specific experiments and the final values of cross sections in those experiments we reach for the EXFOR database~\cite{otuka_towards_2014}.

From the p side, we include, in particular, the data on elastic scattering off $^{48}$Ca by McCamis \etal~\cite{mccamis_elastic_1986}, at the energies of 21, 25, 30, 35, 40, 45, and $48.4 \, \text{MeV}$ and the data on elastic scattering off $^{208}$Pb by Van Oehrs \etal~\cite{van_oers_optical-model_1974}, at the energies of 24.1, 30.3, 35, 45, and $47.3 \, \text{MeV}$.  Moreover, we include the data by Mani \etal~\cite{mani_elastic_1971} on elastic scattering of $49.35 \, \text{MeV}$ protons off three of the targets: $^{90}$Zr, $^{120}$Sn and $^{208}$Pb.  For the $^{90}$Zr target, we further include the elastic scattering data from the measurements by Van der Bijl \etal~\cite{van_der_bijl_high_1983}, at $21.05 \, \text{MeV}$, by De Swiniarski \etal~\cite{de_swiniarski_elastic_1977}, at $30 \, \text{MeV}$, and by Blumberg \etal~\cite{blumberg_polarizations_1966}, at $40 \, \text{MeV}$.  For $^{120}$Sn, we include the scattering data from the measurements by Ridley and Turner at $30.3 \, \text{MeV}$  \cite{ridley_optical_1964}, by Boyd and Greenless at $39.6 \, \text{MeV}$  \cite{boyd_nuclear-matter_1968} and by Fricke \etal\ at $40 \, \text{MeV}$  \cite{fricke_polarization_1967}.

From the n side, we include the data on elastic scattering off $^{48}$Ca by Mueller \etal~\cite{mueller_asymmetry_2011} at $16.8 \, \text{MeV}$.  Moreover, we include the data on elastic scattering off $^{90}$Zr by Bainum \etal~\cite{bainum_isospin_1978}, at $11 \, \text{MeV}$, and by Wang and Rapaport \cite{wang_neutron_1990}, at $24 \, \text{MeV}$.  Further, we include data on elastic scattering off $^{120}$Sn by Rapaport \etal~\cite{rapaport_neutron_1980}, at $11 \, \text{MeV}$ and by Guss \etal~\cite{guss_optical_1989}, at 13.9 and $16.9 \, \text{MeV}$.  For the $^{208}$Pb target, we include the scattering data by Roberts \etal~\cite{roberts_measurement_1991}, at $8 \, \text{MeV}$, by Delaroche \etal~\cite{delaroche_complex_1983}, at $10 \, \text{MeV}$, by Finlay \etal~\cite{finlay_energy_1984}, at 20, 22, and $24 \, \text{MeV}$, by Rapaport \etal~\cite{rapaport_neutron_1978} at $26 \, \text{MeV}$, and finally by DeVito \etal~\cite{devito_neutron_2012} at 30.4 and $40 \, \text{MeV}$.

\subsection{Fits to Data}

\begin{figure}
\centerline{\includegraphics[width=\linewidth]{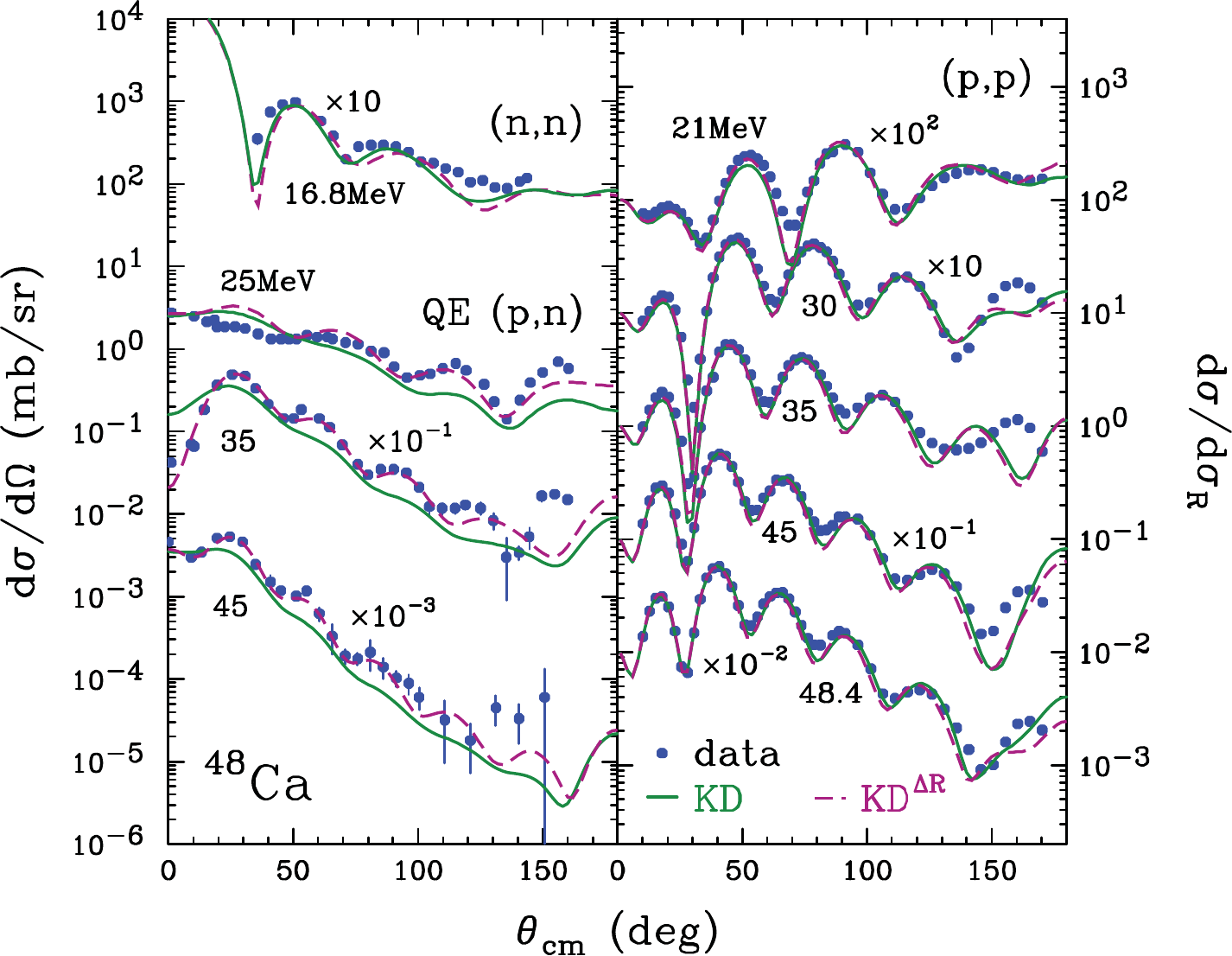}}
\caption{
Differential cross section in elastic and quasielastic reactions on $^{48}$Ca as a function of c.m.\ angle.  The left panel shows differential cross section, in mb/sr, in the reactions with an outgoing neutron, i.e.\ elastic n scattering and quasielastic (p,n) reactions.  The right panel shows differential cross section in elastic p scattering, normalized with the Rutherford cross section.  The~data are represented with filled circles and the theory is represented with lines, solid - for predictions of the original KD optical-potential parametrization \cite{koning_local_2003} and dashed - for the best-fit modified parametrization.  For display purposes, the results for different indicated incident energies are multiplied by different indicated factors.  The (n,n) data are from Ref.~\cite{mueller_asymmetry_2011}, the (p,n) data - from Refs.~\cite{patterson_energy-dependent_1976,doering_microscopic_1975,doering_isobaric_1974} and (p,p) - from Refs.~\cite{mccamis_elastic_1986}.
}
\label{fig:ca48}
\end{figure}

\begin{figure}
\centerline{\includegraphics[width=\linewidth]{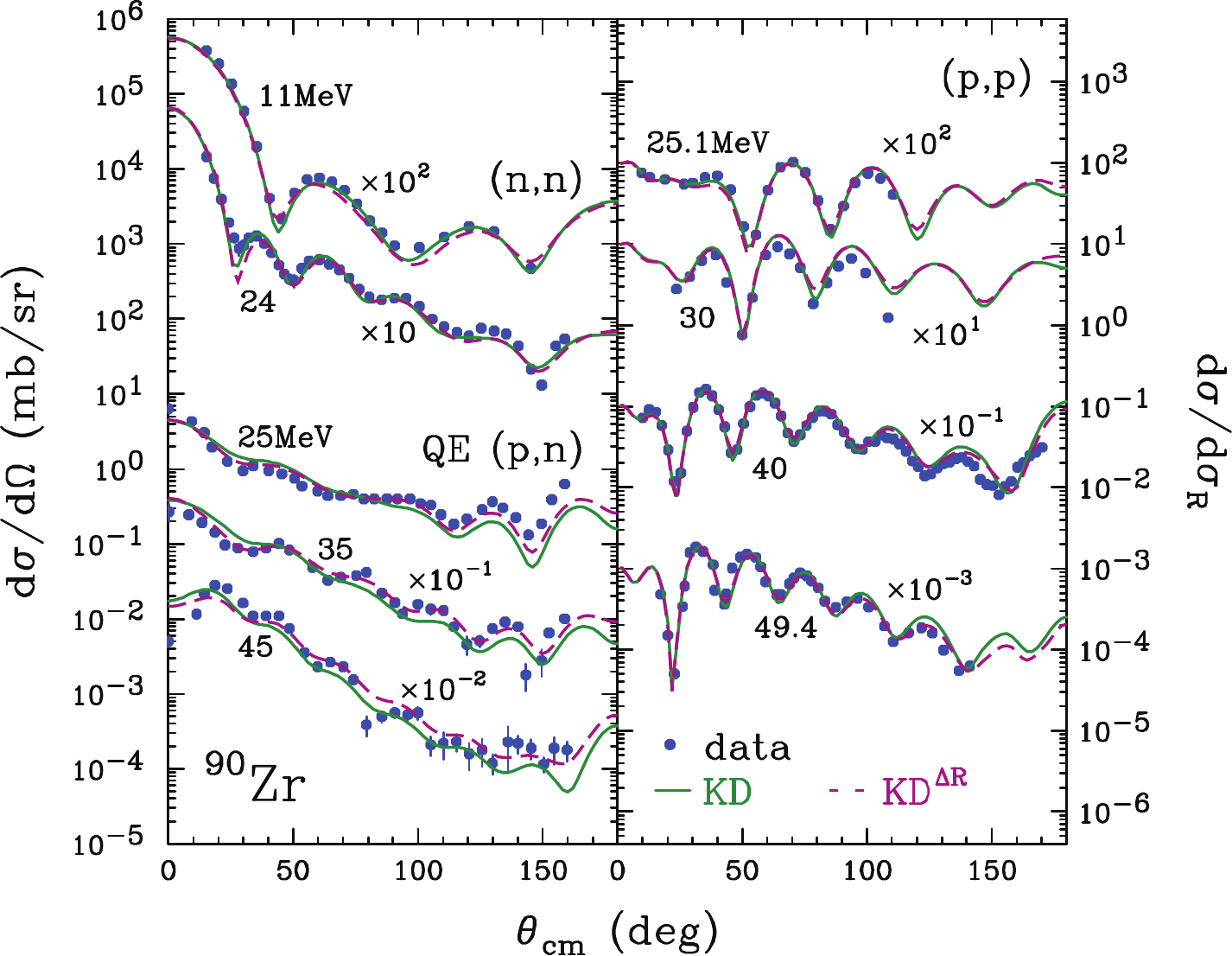}}
\caption{
Differential cross section in elastic and quasielastic reactions on $^{90}$Zr as a function of c.m.\ angle.  The left panel shows differential cross section, in mb/sr, in the reactions with an outgoing neutron, i.e.\ elastic n scattering and quasielastic (p,n) reactions.  The right panel shows differential cross section in elastic p scattering, normalized with the Rutherford cross section.  The~data are represented with filled circles and the theory is represented with lines, solid - for predictions of the original KD optical-potential parametrization \cite{koning_local_2003} and dashed - for the best-fit modified parametrization.  For display purposes, the results for different indicated incident energies are multiplied by different indicated factors.  The (n,n) data are from Refs.~\cite{bainum_isospin_1978,wang_neutron_1990}, the (p,n) data - from Refs.~\cite{patterson_energy-dependent_1976,doering_microscopic_1975,doering_isobaric_1974} and (p,p) - from Refs.~\cite{van_der_bijl_high_1983,de_swiniarski_elastic_1977,blumberg_polarizations_1966,mani_elastic_1971}.
}
\label{fig:zr90}
\end{figure}

\begin{figure}
\centerline{\includegraphics[width=\linewidth]{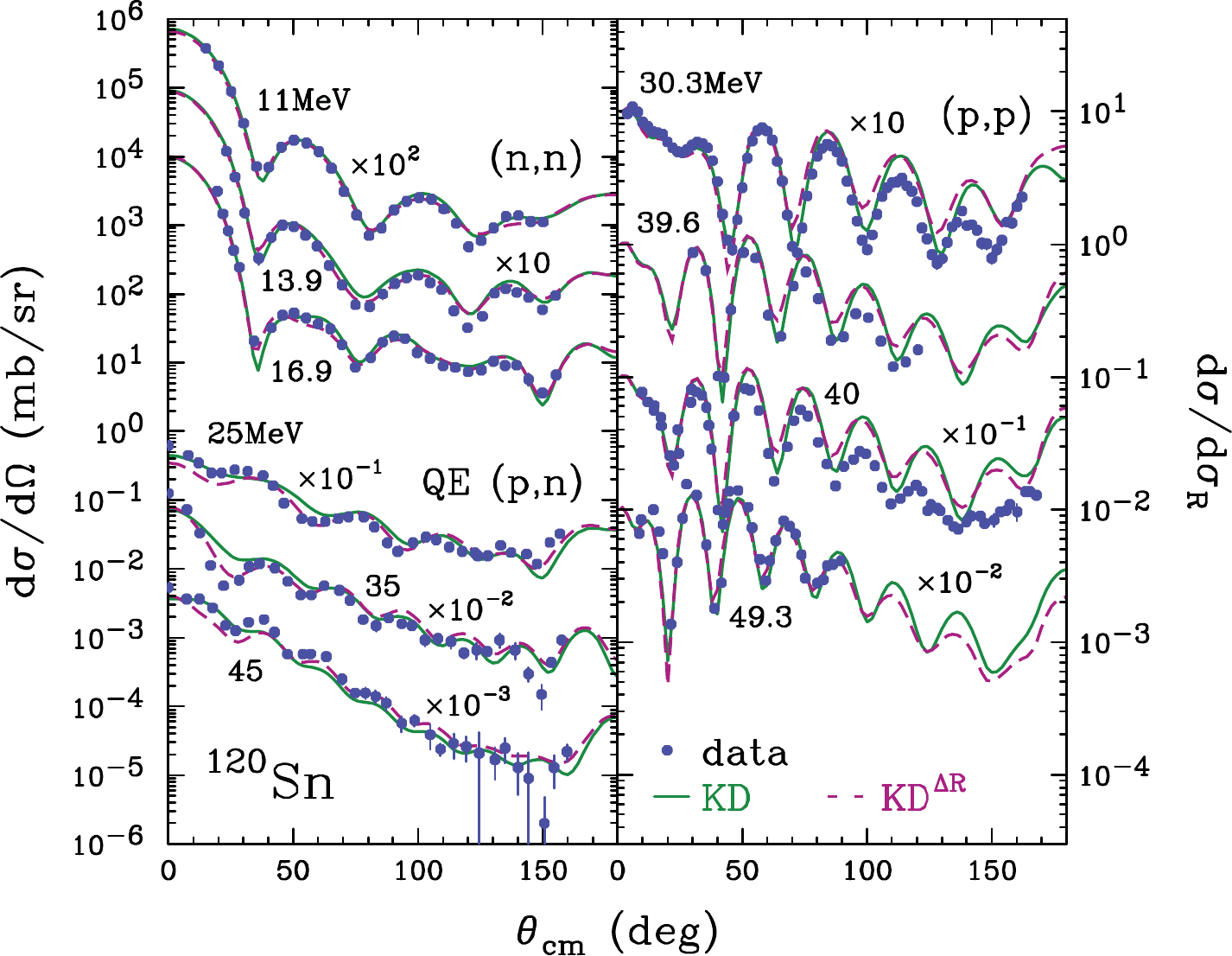}}
\caption{
Differential cross section in elastic and quasielastic reactions on $^{120}$Sn as a function of c.m.\ angle.  The left panel shows differential cross section, in mb/sr, in the reactions with an outgoing neutron, i.e.\ elastic n scattering and quasielastic (p,n) reactions.  The right panel shows differential cross section in elastic p scattering, normalized with the Rutherford cross section.  The~data are represented with filled circles and the theory is represented with lines, solid - for predictions of the original KD optical-potential parametrization \cite{koning_local_2003} and dashed - for the best-fit modified parametrization.  For display purposes, the results for different indicated incident energies are multiplied by different indicated factors.  The (n,n) data are from Refs.~\cite{rapaport_neutron_1980,guss_optical_1989}, the (p,n) data - from Refs.~\cite{patterson_energy-dependent_1976,doering_microscopic_1975,doering_isobaric_1974} and (p,p) - from Refs.~\cite{ridley_optical_1964,boyd_nuclear-matter_1968,fricke_polarization_1967,mani_elastic_1971}.
}
\label{fig:sn120}
\end{figure}

\begin{figure}
\centerline{\includegraphics[width=\linewidth]{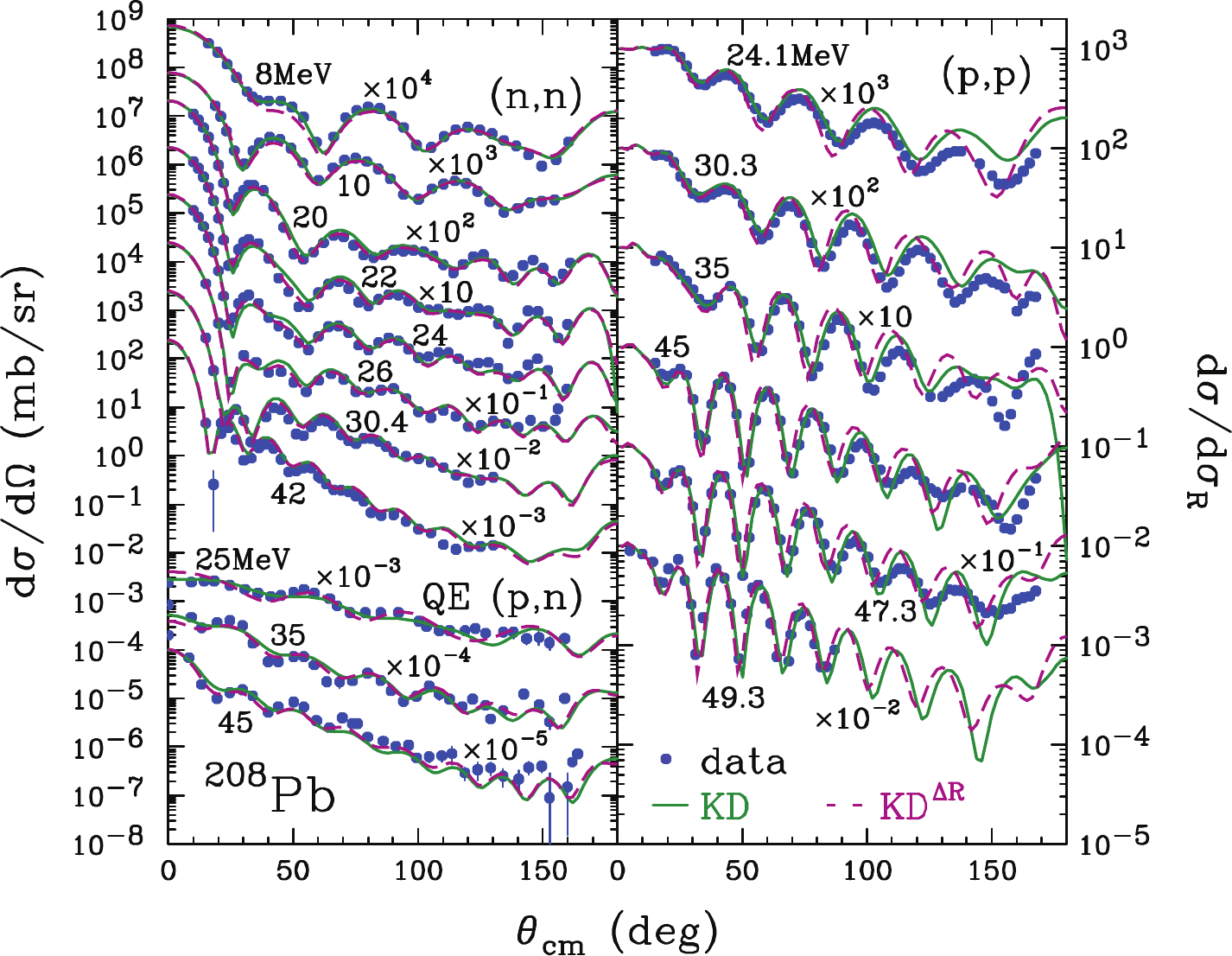}}
\caption{
Differential cross section in elastic and quasielastic reactions on $^{208}$Pb as a function of c.m.\ angle.  The left panel shows differential cross section, in mb/sr, in the reactions with an outgoing neutron, i.e.\ elastic n scattering and quasielastic (p,n) reactions.  The right panel shows differential cross section in elastic p scattering, normalized with the Rutherford cross section.  The~data are represented with filled circles and the theory is represented with lines, solid - for predictions of the original KD optical-potential parametrization \cite{koning_local_2003} and dashed - for the best-fit modified parametrization.  For display purposes, the results for different indicated incident energies are multiplied by different indicated factors.  The (n,n) data are from Refs.~\cite{roberts_measurement_1991,delaroche_complex_1983,finlay_energy_1984,rapaport_neutron_1978,devito_neutron_2012}, the (p,n) data - from Refs.~\cite{patterson_energy-dependent_1976,doering_microscopic_1975,doering_isobaric_1974} and (p,p) - from Refs.~\cite{van_oers_optical-model_1974,mani_elastic_1971}.
}
\label{fig:pb208}
\end{figure}

Figures \ref{fig:ca48}-\ref{fig:pb208} display the data discussed above, as filled circles, together with the predictions of the KD parametrization, as solid lines.  It may be seen in these figures that the general description of both the elastic and QE cross sections in terms of the KD parametrization is quite good.  Given that the KD potential parameters have been fitted to describe different elastic data, the quality in describing such data might not be surprising.  However, the fact the quasielastic data, never considered in arriving at the KD parameter values, are also reasonably well described supports both the relevance of the isospin formalism in the Lane potential \eqref{eq:Lane} and the physics validity of the KD parametrization.

When comparing details of the (p,n) predictions of the KD parametrization with data, though, one may notice that the predictions generally exhibit less oscillation with angle than do data.  Given the discussion above, there are reasons to investigate whether this could be attributed to different geometric features of the isovector than isoscalar potentials - in the KD parametrization those features are practically identical.  We progress fitting the parameters representing adjustments in the geometry and overall strength as discussed in Subsection \ref{ssec:strategy}.

In the context of estimating parameter values and their errors with a $\chi^2$ minimization, the following should be stated.  The experimental errors on p elastic scattering cross sections can be quite small.  The model relying on a simple potential form, with few parameters, has no chance of describing the elastic data with an arbitrary precision, i.e.\ yield $\chi^2$ values per degree of freedom (DOF) of the order of 1 no matter what the experimental errors are.  It can be seen in Figs.~\ref{fig:ca48}-\ref{fig:pb208} that, in the rear direction, the model can struggle even at qualitative level, not just quantitative.  When the cross sections from direct amplitudes are low, other processes can effectively compete.  One of the standard strategies of statistical analysis in such a~situation is that of $\chi^2$ renormalization, effectively recognizing that an error residue represents limitations of a model that cannot be circumvented.  If the renormalization is done, though, only after a $\chi^2$ value constructed with experimental errors gets minimized, the best-fit theoretical cross sections may exhibit awkward features.  Namely the portions of cross sections where the experimental errors are very small may get reproduced very well within the fit, while the remainder may be left largely abandoned.  E.g.\ if elastic proton and neutron elastic cross sections are combined in $\chi^2$ with experimental errors, the neutron cross sections may bear nearly no impact on the outcome of the fit.  We find it more effective to assume some systematic theoretical errors from scratch in the fit, of magnitude that would yield a final $\chi^2/\text{DOF}$ of the order of few.  Afterwards still the standard $\chi^2$ renormalization may be applied.  In this way, the regions with larger experimental errors, that may be still fairly small on the scale of accuracy expected from the theory, may impact the fit as much as the regions with very small experimental errors, yielding a more democratic fit.

In our case we find that assumed theoretical errors of the order of (5--10)\% yield cross-section fits that are visually acceptable.  Given the large experimental errors for the (p,n) reaction cross sections, the inflating of the errors ends up with quite limited impact on the parameters characterizing the relation between isoscalar and isovector potentials, i.e.\ $\Delta R$, $\Delta a$ and ${\mathcal F}$, and their errors.  However, the impact is more significant on the parameters fine-tuning the characteristics of the isoscalar potentials, i.e.\ $\Delta R_0$, $\Delta a_0$ and ${\mathcal F}_0$, with the parameters values generally drifting either towards a better reproduction of (p,p) cross sections, for lower assumed theoretical errors, or towards a better reproduction of (n,n) cross sections, for higher errors.  When quoting errors on the best-fit parameter values, we include there the specific drift under changing assumptions on the theoretical errors.

\begin{table}
\caption{Adjustment parameters for the isoscalar and isovector components of nucleonic potentials constructed from the KD parametrization \cite{koning_local_2003}, representing the best simultaneous description of elastic (p,p) and (n,n) and quasielastic (p,n) reactions on different targets, cf.\ Figs.~\ref{fig:ca48}--\ref{fig:pb208}.  The~adjustment parameters for the isoscalar potential, ${\mathcal F}_0$, $\Delta R_0$ and $\Delta a_0$, are relative to the modified ($a_D = 0.570 \, \text{fm}$) KD parametrization.  The adjustment parameters pertaining to the isovector potential, ${\mathcal F}$, $\Delta R$ and~$\Delta a$, are {\em relative to the adjustments in the isoscalar potential}, i.e.~e.g.~$\Delta R_1=\Delta R_0 + \Delta R$.\\[-2.5ex]}
\label{tab:BestFit}
\begin{tabular}{||c||c|c|c||c|c|c||}
\hline
\hline
Target & ${\mathcal F}_0$ & $\Delta R_0$ & $\Delta a_0$  & ${\mathcal F}$ & $\Delta R$ & $\Delta a$ \\
       &                  &  [fm]        &  [fm]         &                  &  [fm]        &  [fm]   \\
\hline
\hline
$^{48}$Ca &  0.978$\, \pm \,$0.008  & -0.002$\, \pm \,$0.012 & -0.037$\, \pm \,$0.004 & 0.835$\, \pm \,$0.038  & 0.84$\, \pm \,$0.08 &  \\
$^{90}$Zr &  1.045$\, \pm \,$0.011  & -0.130$\, \pm \,$0.034 & 0.019$\, \pm \,$0.016 & 0.814$\, \pm \,$0.030 & 0.87$\, \pm \,$0.09 &  \\
 $^{120}$Sn & 1.031$\, \pm \,$0.008  & -0.101$\, \pm \,$0.021 & 0.038$\, \pm \,$0.009 & 0.689$\, \pm \,$0.047 & 1.14$\, \pm \,$0.18  &  \\
$^{208}$Pb &  1.010$\, \pm \,$0.008 & 0.032$\, \pm \,$0.034 & 0.034$\, \pm \,$0.016 & 0.799$\, \pm \,$0.028 & 1.08$\, \pm \,$0.17 &  \\
\hline
    All     &   &  &  &  &  &  -0.104$\, \pm \,$0.033  \\
\hline
\hline
\end{tabular}
\end{table}

Optimal parameter values, when adjusting ${\mathcal F}_0$, $\Delta R_0$, $\Delta a_0$, ${\mathcal F}$ and $\Delta R$ on per-isobar basis, and $\Delta a$ globally, are listed in Table \ref{tab:BestFit}.  The corresponding differential cross sections are represented with dashed lines in Figs.~\ref{fig:ca48}--\ref{fig:pb208}.  While the description of the (p,n) data generally significantly improves across nuclei, compared to the original KD parametrization, this may come at the cost of some deterioration in the description of (p,p) and (n,n) data.

\begin{figure}
\centerline{\includegraphics[width=.65\linewidth]{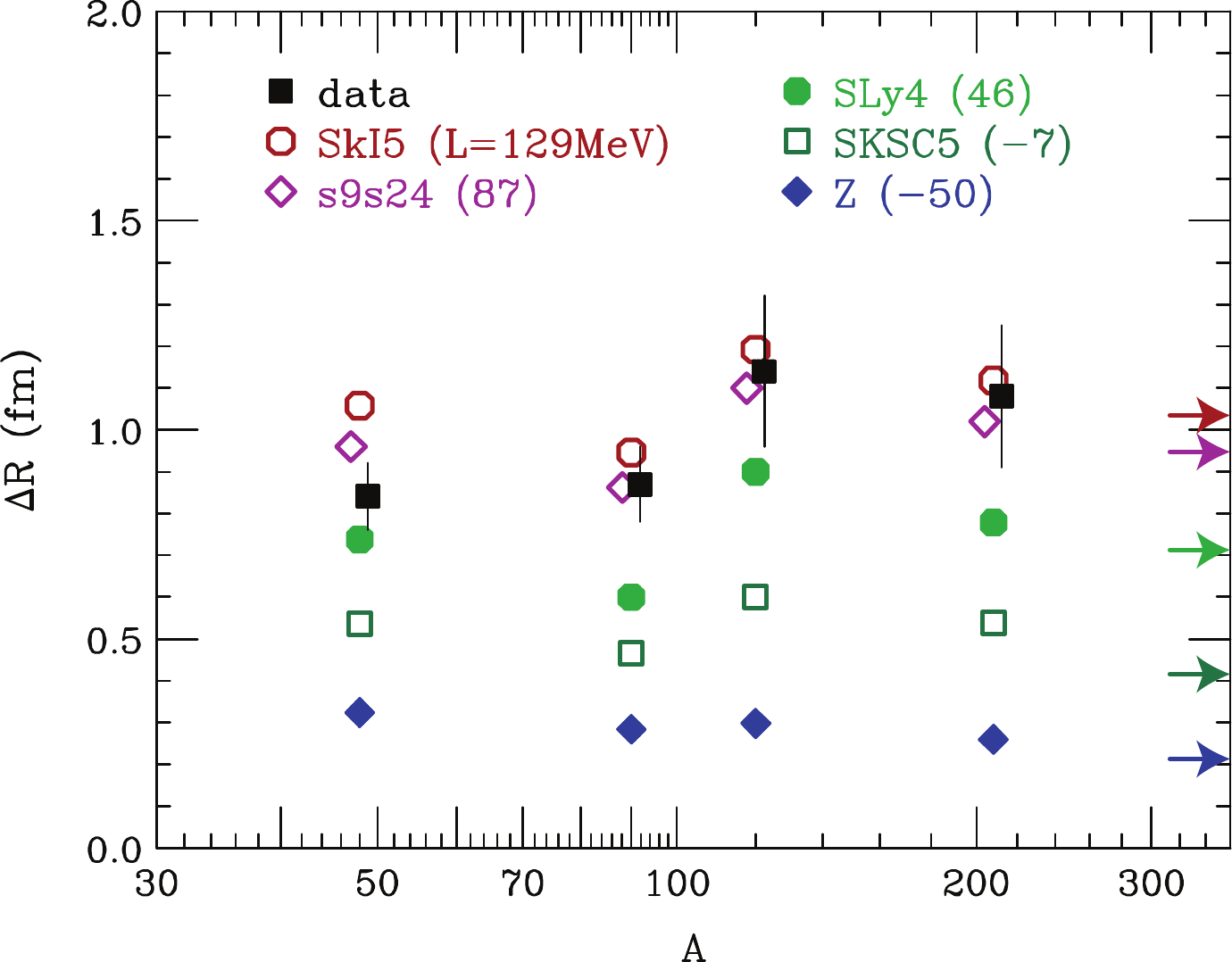}}
\caption{
Displacement of isovector surface, relative to isoscalar, displayed vs mass number.  Squares represent the displacement values arrived at when simultaneously adjusting isovector and isoscalar potential parameters, in the modified KD potential parametrization, to best describe the measured differential quasielastic (p,n) and elastic (p,p) and (n,n) cross sections on the same targets.  The~remaining symbols represent the displacement of isovector surface relative to isoscalar arrived at for densities from SHF calculations with different indicated Skyrme parametrizations.  In the parenthesis in the figure, values of the associated slope parameters $L$ of the symmetry energy are given in MeV, following the names of the Skyrme parametrizations.  Arrows by the right axis indicate displacement values in the limit of half-infinite matter, after~I, for the specific parametrizations in the order of decreasing $L$-values, from top to bottom.
}
\label{fig:drx}
\end{figure}

\begin{figure}
\centerline{\includegraphics[width=.65\linewidth]{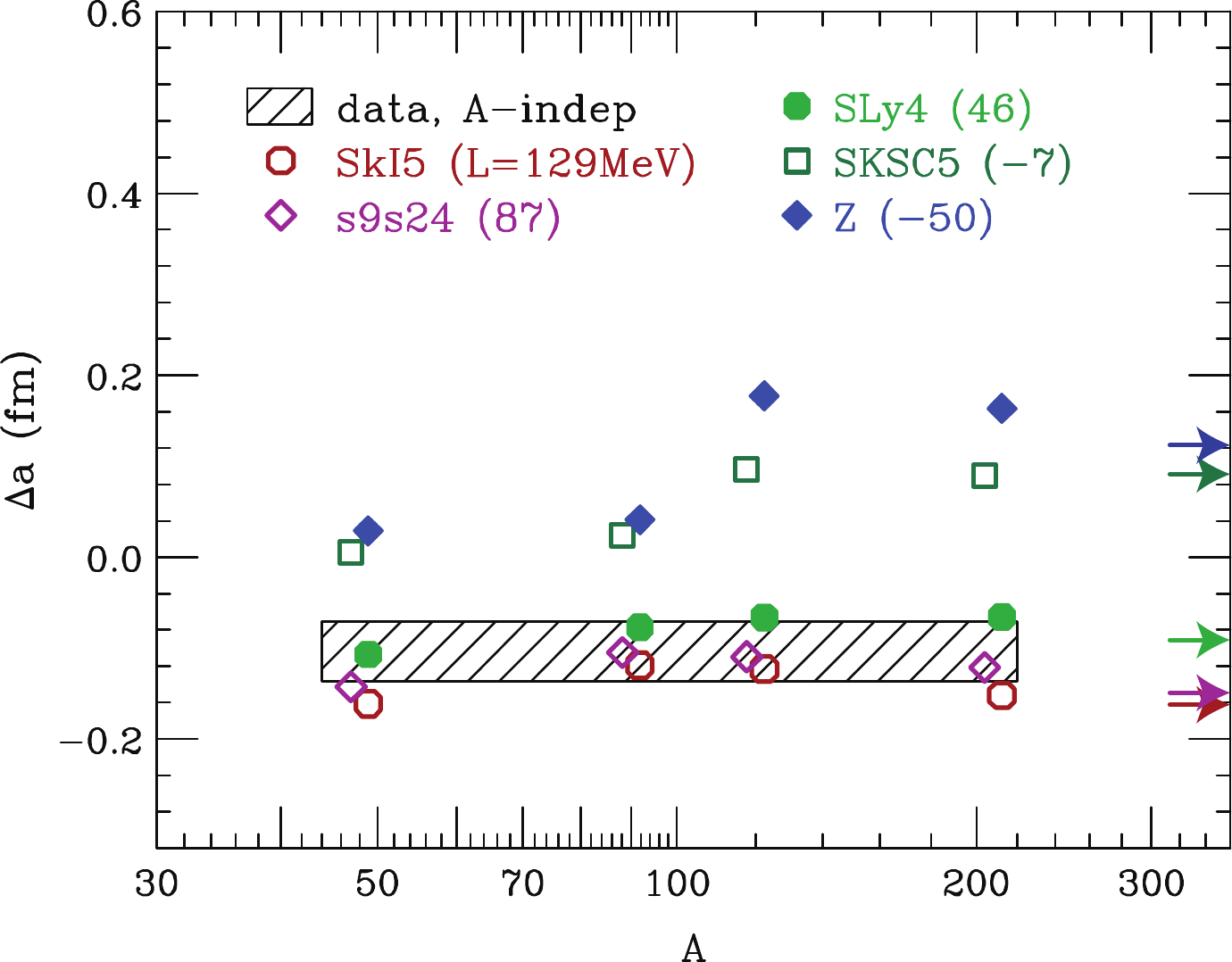}}
\caption{
Difference between diffusivities of isovector and isoscalar surfaces, displayed vs mass number.  The shaded region represents a mass-independent difference between diffusivities of potential surfaces arrived at when simultaneously adjusting different isovector and isoscalar potential parameters, in the modified KD potential parametrization, to best describe the measured differential quasielastic (p,n) and elastic (p,p) and (n,n) cross sections on the same targets.  The~remaining symbols represent the difference between diffusivities of the isovector and isoscalar surfaces arrived at for densities from SHF calculations with different indicated Skyrme parametrizations.  In the parenthesis in the figure, values of the associated slope parameters $L$ of the symmetry energy are given in MeV, following the names of the Skyrme parametrizations.  Arrows by the right axis indicate difference in diffusivity for the surfaces in the limit of half-infinite matter, after~I, for the specific parametrizations in the order of increasing $L$-values, from top to bottom.
}
\label{fig:dax}
\end{figure}

It may be seen in Table \ref{tab:BestFit} that the best-fit adjustments for the isoscalar potential are relatively minor, i.e.\ ${\mathcal F}_0$ tends to be close to 1 and $\Delta R_0$ and $\Delta a_0$ are small.  Moderate evolution in $\Delta a_0$ from $^{48}$Ca to heavier nuclei mimicks the growth of $a_D$ for protons in the KD parametrization~\cite{koning_local_2003}.  The prominent adjustments in the Table are those for the isovector potential, in favored weakening of the strength by $\sim 20 \%$ and pushing out of the surface, relative to isoscalar, by $\sim 1 \, \text{fm}$.  The fit across all isobars favors also some steepening of the isovector surface relative to isoscalar, with relative drop in the slope parameters by $\sim 0.1 \, \text{fm}$.  Besides the Table, we represent the deduced values of $\Delta R$ and $\Delta a$ in Figs.~\ref{fig:drx} and~\ref{fig:dax}.

If we force the value of $\Delta R$ to be independent of $A$, we arrive at the optimal parameter values listed in Table~\ref{tab:BestFit1}.  The assumption of universal surface displacement, yields $\Delta R = 0.88 \pm 0.05 \, \text{fm}$ and $\Delta a = -0.093 \pm 0.024 \, \text{fm}$.  With parameters in Table \ref{tab:BestFit1} close those in Table \ref{tab:BestFit}, the description of differential cross sections is very similar to that in Figs.~\ref{fig:ca48}--\ref{fig:pb208}.  With the by-eye differences relative to Figs.~\ref{fig:ca48}--\ref{fig:pb208} being hardly detectable, we refrain from producing another set of figures tied to Table \ref{tab:BestFit1}.

\begin{table}
\caption{Adjustment parameters for the isoscalar and isovector components of nucleonic potentials constructed from the KD parametrization \cite{koning_local_2003}, representing the best simultaneous description of elastic (p,p) and (n,n) and quasielastic (p,n) reactions on different targets, when forcing the displacement of isovector surface relative to isoscalar to be $A$-independent.   The~listed adjustment parameters for the isoscalar potential, ${\mathcal F}_0$, $\Delta R_0$ and $\Delta a_0$, are relative to the modified ($a_D = 0.570 \, \text{fm}$) KD parametrization.  The adjustment parameters pertaining to the isovector potential, ${\mathcal F}$, $\Delta R$ and~$\Delta a$, are {\em relative to the adjustments of the isoscalar potential}.\\[-2.5ex]}
\label{tab:BestFit1}
\begin{tabular}{||c||c|c|c||c|c|c||}
\hline
\hline
Target & ${\mathcal F}_0$ & $\Delta R_0$ & $\Delta a_0$  & ${\mathcal F}$ & $\Delta R$ & $\Delta a$ \\
       &                  &  [fm]        &  [fm]         &                  &  [fm]        &  [fm]   \\
\hline
\hline
$^{48}$Ca &  0.976$\, \pm \,$0.006  & 0.003$\, \pm \,$0.007 & -0.041$\, \pm \,$0.005 & 0.808$\, \pm \,$0.035  & &  \\
$^{90}$Zr &  1.048$\, \pm \,$0.010  & -0.150$\, \pm \,$0.028 & 0.023$\, \pm \,$0.010 & 0.795$\, \pm \,$0.032 & &  \\
 $^{120}$Sn & 1.036$\, \pm \,$0.007  & -0.115$\, \pm \,$0.014 & 0.027$\, \pm \,$0.005 & 0.738$\, \pm \,$0.031 & &  \\
$^{208}$Pb &  1.012$\, \pm \,$0.006 & 0.016$\, \pm \,$0.027 & 0.024$\, \pm \,$0.010 & 0.826$\, \pm \,$0.035 & &  \\
\hline
    All     &   &  &  &  & 0.88$\, \pm \,$0.05  &  -0.093$\, \pm \,$0.024  \\
\hline
\hline
\end{tabular}
\end{table}

\section{Densities in Skyrme-Hartree-Fock Calculations}

\subsection{Isoscalar and Isovector Densities}

We discussed the isoscalar and isovector densities for half-infinite nuclear matter within SHF  in I.  In II, we used the densities from spherical SHF calculations, to arrive at symmetry-energy coefficients for individual nuclei.  The latter coefficients can be expressed in terms of an integral of density of transverse isospin in a nucleus (termed also in I and II as asymmetric density).  Coulomb interactions polarize the difference of neutron and proton densities making the density profile of the third component different from the profile of transverse components.  While direct extraction of the difference of neutron and proton densities, $\rho_3 = \rho_n - \rho_p$, is obviously trivial in SHF, similar extraction of density for the transverse components is not, due to HF breaking isospin invariance.  Fortunately, as indicated in II, owing to slow changes in Coulomb potential, compared to nuclear nonlocalities, in the bulk low-asymmetry limit, the density tied to transverse isospin can be obtained from the density of the third component through a renormalization with the local chemical potential for asymmetry:
\beq
\rho_\perp(r) \propto \frac{\rho_3(r)}{\mu_3 + \Phi(r)/2} \, .
\label{eq:rhoT}
\eeq
Here, $\mu_3$ is the global potential for asymmetry,
\beq
\mu_3 = \frac{\partial E}{ \partial (N-Z)} \bigg|_A \, ,
\eeq
and $\Phi$ is the local Coulomb potential for protons.  The chemical potential itself can be arrived
at by using the discussed densities, cf.~II.
Besides the Coulomb field, the isovector densities are very sensitive to shell effects, far more than the net densities, because of their differential nature, as evident already in half-infinite matter in I, where the role of shell effects was taken over by the Friedel oscillations.  Incidentally, the very use of the densities in assessing the chemical potential, and integrating then over isovector density, aims at reducing the impact of the shell effects on the potential as compared to direct energy differentials.

Commenting more, the renormalization \eqref{eq:rhoT} aims to produce the isospin distribution $\rho_\perp(r)$ without the effects of a~Coulomb polarization - at the microscopic level one for a pure state with isospin~$T$.  The~distribution $\rho_3 (r)$ is for a state where, microscopically, the Coulomb interactions admix the states that differ in isospin from $T$, primarily the $T'=T+1$ state where an isospin-0 core is replaced by an isospin-1 giant monopole oscillation~\cite{lane_isobaric_purity_1962,auerbach_coulomb_1983}.  Within e.g.\ $^{48}$Ca, that core may be perceived as $^{40}$Ca that gets polarized and for which the Coulomb-induced isovector modulation may be approximated in terms of the amplitude for a monopole oscillation.  The excess neutrons, in the pure state~$T$ outside of the core in this microscopic transcription, distribute themselves according to $\rho_\perp(r)$, while the overall neutron-proton imbalance according to~$\rho_3(r)$.

\begin{figure}
\centerline{\includegraphics[width=\linewidth]{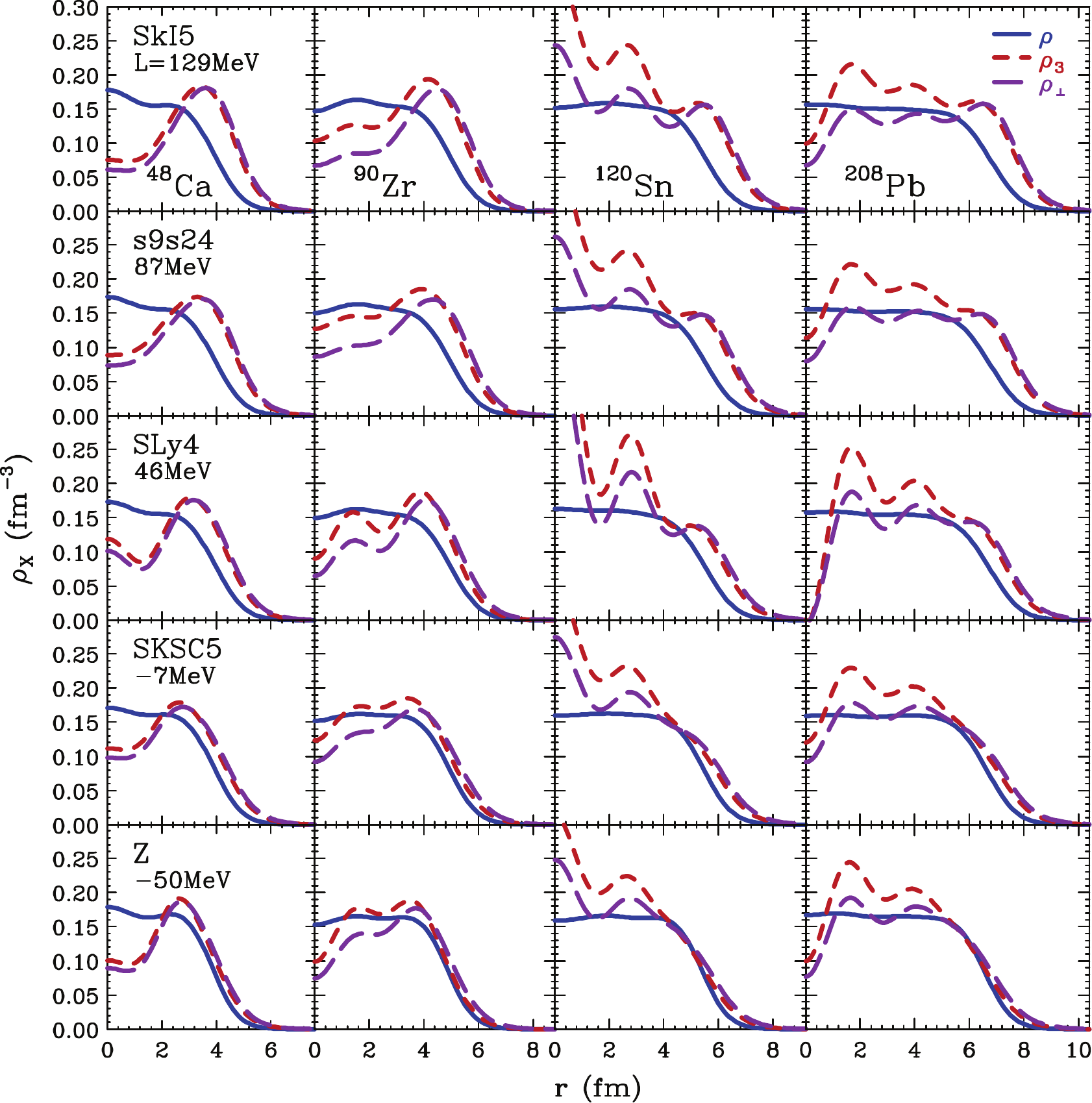}}
\caption{
Densities from spherical SHF calculations with different Skyrme interaction parameterizations for different nuclei.  For the panels from top to bottom the parametrizations are SkI5~\cite{1995NuPhA.584..467R}, s9s24~\cite{brown_constraints_2013}, SLy4~\cite{1998NuPhA.635..231C}, SKSC5~\cite{PhysRevC.50.460} and Z~\cite{PhysRevC.33.335}.  Underneath the interaction names in the figure, values of the slope of the symmetry energy with respect to density are given.  For the panels from left to right the nuclei are $^{48}$Ca, $^{90}$Zr, $^{120}$Sn and $^{208}$Pb.  The net density $\rho$ (solid lines) is presented in absolute normalization.  The isovector density $\rho_\perp$ (longer-dashed lines) is normalized to the same average density in the interior as $\rho$.  The isovector density $\rho_3$ (shorter-dashed) is normalized to the same average density as $\rho_\perp$ in the region immediately adjacent to the surface.
}
\label{fig:rho_all}
\end{figure}

Figure \ref{fig:rho_all}
exhibits densities arrived at from SHF calculations with the code of P.-G.~Reinhard~\cite{Reinhard:1991}.  Different columns of panels there represent different nuclei and different rows represent different interaction parametrizations.  In Fig.\ref{fig:rho_all}, the net density~$\rho$ is represented in its absolute normalization, with solid lines.  The densities $\rho_3$ and $\rho_\perp$ are represented, respectively, with shorter- and longer-dashed lines.  The density $\rho_\perp$ is normalized to the same average density in the interior as $\rho$.  The density $\rho_3$ is normalized to the same average density as $\rho_\perp$, within the interior region immediately adjacent to the surface.  For the average density in the normalizations, we first assess the position of the surface for the given density and then count, as the interior for that density, the volume with radius shorter by $0.8 \, \text{fm}$.  As the interior immediately adjacent to the surface, we count the outer layer of the interior volume having $1 \, \text{fm}$ thickness.
The density normalizations are tied to the characteristics and use of those densities.  The values of $\rho$ and $\rho_\perp$ are expected to stabilize within the interior of a large system, hence it makes sense to normalize $\rho_\perp$ to the same average value in the interior as $\rho$, when emphasizing any similarities and differences in the shape between the two.  In the case of $\rho_3$, we want to emphasize that it is going to yield similar results as $\rho_\perp$ when used exclusively in the surface region, hence the normalization to the same average density as $\rho_\perp$ in the region immediately adjacent to the surface.

In examining the panels from left to right in Fig.~\ref{fig:rho_all}, one can see an evolution of the densities with system size.  In examining the panels from top to bottom one can see the evolution of densities with changing slope $L$ of the symmetry energy with respect to density.  From left to right in Fig.~\ref{fig:rho_all}, the strengthening of the Coulomb effects is evident in an~increasing departure of $\rho_3$ from~$\rho_\perp$, and in this $\rho_3$ growing on the average towards the nuclear center.   For the $^{48}$Ca nucleus, with weak Coulomb effects, those two densities, $\rho_3$~and~$\rho_\perp$, tend to be nearly the same, no matter what the nuclear interaction.  In Fig.~\ref{fig:rho_all}  it is further apparent that the shell effects are much more pronounced in the isovector than in the isoscalar densities.  The shell effects in density can be described as oscillations at about half of the Fermi wavelength.  The wavelength is a bit longer for the lighter than heavier systems.  Irrespectively of the shell effects giving rise to differences between the nuclei, due to the different number of wavelengths fitting into the size, the general expectation is that the surface in any nucleus synchronizes the rise in the particle wavefunctions towards the interior and the positions of the first nodes.  There is a difference in the impact on neutrons and protons at finite asymmetry, due to a difference in the Fermi energy of the particles and due to any interaction differences tied to the symmetry energy.  Because of the synchronization, the isovector densities may be expected to have a~universal appearance and a~universal relation to the isoscalar density in the surface region, no matter what the mass of a nucleus.  However, the surface also has a different impact on the neutrons and protons due to the Coulomb interactions.  To the latter impact can be attributed the gradual drift for $\rho_3$ in relation to $\rho$ in the surface region, that rises with charge number in Fig.~\ref{fig:rho_all}, from the advocated universality rule.  After $\rho_3$ gets renormalized to yield $\rho_\perp$, though, we can observe that, for a~given interaction the same relation between $\rho_\perp$ and $\rho$ generally persists, no matter what the nuclear mass.  This, in particular, underscores the significance of the half-infinite matter considerations such as in~I.  It should be mentioned, though, that while the density $\rho_\perp$ should be a unique functional of $\rho$ in the bulk low-asymmetry limit, cf.~II, the Coulomb interactions have an impact on $\rho$.  That impact is not removed in~\eqref{eq:rhoT}, while the Coulomb interactions are switched off with all their impact for the half-infinite matter.

In examining the evolution of the characteristics of the densities with changes in the symmetry energy in Fig.~\ref{fig:rho_all}, i.e.\ the panels top to down, we can see that the isovector and isoscalar surfaces approach each other as $L$ decreases.  For high $L$, the $\rho_\perp$-surface is steeper than the $\rho$-surface, but for low $L$ the situation is generally reversed.  Such findings where first arrived at within the examinations of half-infinite matter in~I.
  It should be noted that these general characteristics of evolution with $L$ become evident for sufficiently large changes in~$L$. For smaller changes in $L$, the evolution in any particular characteristics of the densities, especially in a finite system, might not appear monotonic; obviously various characteristics of the densities depend also on aspects of the interactions that have no direct relation to the symmetry energy.

\subsection{Comparison to Data}

In comparing the theory to data, we compare the geometric relation of $\rho_\perp$ to $\rho$ to the corresponding relation of $U_1$ to $U_0$, where $U_1$ is predominantly determined from the quasielastic charge exchange reactions.  In a more sophisticated analysis one might ponder making~$U_1$ anisotropic in isospin space, with the $U_3$ component tied to $\rho_3$ and $U_\perp$ to $\rho_\perp$.  While $\rho_3$ is not very different from $\rho_\perp$ in shape, when the two are considered exclusively in the surface region, cf.~Fig.~\ref{fig:rho_all}, we want to stress that it is $\rho_\perp$, rather than $\rho_3$, that is tied to the charge-exchange reactions leading to IAS.  The similarity of the densities there is due to the fact that the Coulomb potential is not varying quickly enough across the surface region to make a difference even for heavy nuclei.  Arguments for tying $\rho_\perp$ to the QE charge exchange include those from macroscopic and microscopic side.  In classical considerations of isospin within an isobaric multiplet, the transverse components of isospin are distributed according to $\rho_\perp$ as already indicated earlier.  When viewed from the microscopic side, that density represents transition density between adjacent members of the multiplet.  From the microscopic side, when directly examining a matrix element of transverse isospin operator between a ground state and its isobaric analog, there will be a contribution from overlap of pure isotopic states, to be accounted for by $\rho_\perp$.  Additional contribution might come from conversion of admixtures with different isospin to one state to those of the other. However, no good match can be achieved when changing the state of only one nucleon with the isospin operator and that contribution will be suppressed.  Last but not least, the fit to the QE charge-exchange data gives no indication for a drop in isovector skin size with increase in target mass, suggested by Fig.~\ref{fig:rho_all} if $\rho_3$ were used.  As one more remark concerning the densities, the renormalization~\eqref{eq:rhoT} can restore average properties of $\rho_\perp$ from $\rho_3$, but likely not the details of shell effects.  Within the focus on the surface region, this should not be an issue, though.

For the sake of comparisons to data, in the combination of the net density $\rho$ for a nucleus and the isovector density $\rho_\perp$ normalized as in Fig.~\ref{fig:rho_all}, we determine the separation of the surfaces by finding the separation $\Delta R$ between the locations where these densities fall to the same value $\rho(0)/2$.  We determine the diffusivity~$a_X$ for the density $\rho_X$, at the corresponding location above, from
\beq
a_X = \frac{\rho_X}{2 \, \text{d}\rho_X/\text{d} r} \, ,
\eeq
guided by a WS shape.  Under the assumption that any significant differences in geometry for the isoscalar and isovector potentials explored in the direct reactions will primarily follow the differences in the geometry for the isoscalar and isovector densities, we plot in Figs.~\ref{fig:drx} and~\ref{fig:dax} the separations between the surfaces and differences in the diffusivity for the cases of the interactions in Fig.~\ref{fig:rho_all}.  In addition to the results for finite nuclei, we provide in Figs.~\ref{fig:drx} and~\ref{fig:dax} the reference results for half-infinite matter, following I, represented in these figures with arrows by the right axes.


\section{Symmetry Energy and Neutron Matter}

\subsection{General Strategy}

As the values of $\Delta R$ and $\Delta a$ appear correlated with the slope of symmetry energy $L$ in the SHF calculations, cf.\ Figs.~\ref{fig:drx} and~\ref{fig:dax}, we attempt to use the values of $\Delta R$ and $\Delta a$ inferred from the reaction fits to constrain $L$ in the structure calculations.  As discussed in~I, an even tighter correlation of $\Delta R$ is expected with $L/a_a^V$, than with $L$ alone.  Here $a_a^V$ is the value of the symmetry energy at normal density, $a_a^V \equiv S(\rho_0)$.  On its own, the geometric quantities poorly constrain the strength of the symmetry energy, though.  To constrain both $L$ and $a_a^V$, the best strategy might be then to combine the conclusions following from $\Delta R$ or~$\Delta a$, with those following from the magnitude of symmetry coefficients for individual nuclei, such as deduced in II from the excitation energies to ground-state IAS.  In II we combined the conclusions from the latter excitation energies with the conclusions following from the sizes of neutron skins~$\Delta r$.  The accuracy of the experimental values for the latter has been a~source of concern - results from different sources in the literature disagreed with each other by more than expected on the basis of the claimed errors for those results.  The vast majority of the results for $\Delta r$ stemmed from analysis of elastic scattering data for which the sensitivity of cross sections to skins is of the type such as illustrated in the top panel of Fig.~\ref{fig:ppn48Ca35_fvr}.

To simultaneously constrain $L$ and $a_a^V$, we progress then in parallel to II, using the symmetry coefficients $a_a(A)$ arrived at in II, proportional to the excitation energies to IAS, $E^*_\text{IAS}$, in combination with the geometric characteristics tied to the symmetry energy.
In II, as the geometric characteristics we utilized $\Delta r$ for different nuclei and here we utilize $\Delta R$ and $\Delta a$.  Given the scarcity of Skyrme parametrizations that need to meet requirements, when many requirements are imposed, we assume that it is possible to find a path in the parameter space, between any two parametrizations that are close enough in their predictions, such that the corresponding predictions vary in a linear fashion with location along the path.  With this we linearly interpolate and moderately (33\% in both directions) extrapolate physical predictions and other characteristics between and around the established interaction parametrizations that are close in their predictions.  Similar strategies are employed in statistical analysis elsewhere, see e.g.~\cite{sangaline_toward_2016}, typically to reduce computational effort.  The~interpolations allow, in particular, to explore better the boundaries of a constraint region in $(L,a_a^V)$ plane, when the number of parametrizations meeting the requirements is low.

In II, to arrive at specific conclusions on symmetry energy, we applied gates of consistency with data, within errors, to the Skyrme parameterisations.  Here, instead, we employ Bayesian inference~\cite{dagostini_bayesian_2003}, constructing density of probability in the space of explored quantities, first uniform within the range of $(L,a_a^V)$ that is spanned by the Skyrme parametrizations in the literature.  The Skyrme parametrizations and their mentioned combinations span for us possible connections between theoretical inputs and predictions.  With inclusion of value~$\overline{E}$ for observable~$E$, determined with experimental error $\sigma_E$, the density of pobability in quantity~$x$ is updated according to
\beq
p(x|\overline{E}) \propto p_\text{prior}(x)  \int \text{d}E \, \text{e}^{-\frac{(E-\overline{E})^2}{2 \sigma_E^2}} \, p(E|x) \, .
\eeq
Here, $p(a|b)$ stands for the conditional probability of $a$ subject to $b$ being true, with $\int \text{d}a \, p(a|b) = 1 $, for mutually exclusive and exhaustive occurances $a$.  Specifically, $p(E|x)$ stands for the probability of arriving at a value $E$ of an observable for theoretical inputs~$x$, excluding yet impact of the uncertainty in a specific measurement.  Typical theory activities give rise to such probabilities and, as mentioned before, we use the different Skyrme parametrizations and their combinations with symmetry energy characterized by similar $(L,a_a^V)$, to arrive at the span of observable values representing any specific $(L,a_a^V)$-region.  After accounting for data on $E$, the resulting probability density $p(x|\overline{E})$ may be used as a prior density, when accounting for any subsequent data.  In the limit of large number of data employed to constrain the probability, it is expected to be only weakly dependent on the assumed original prior.  A~stronger dependence of the posterior on the prior is expected for limited data.

\subsection{Characteristics of Uniform Matter}

\begin{figure}
\centerline{\includegraphics[width=.7\linewidth]{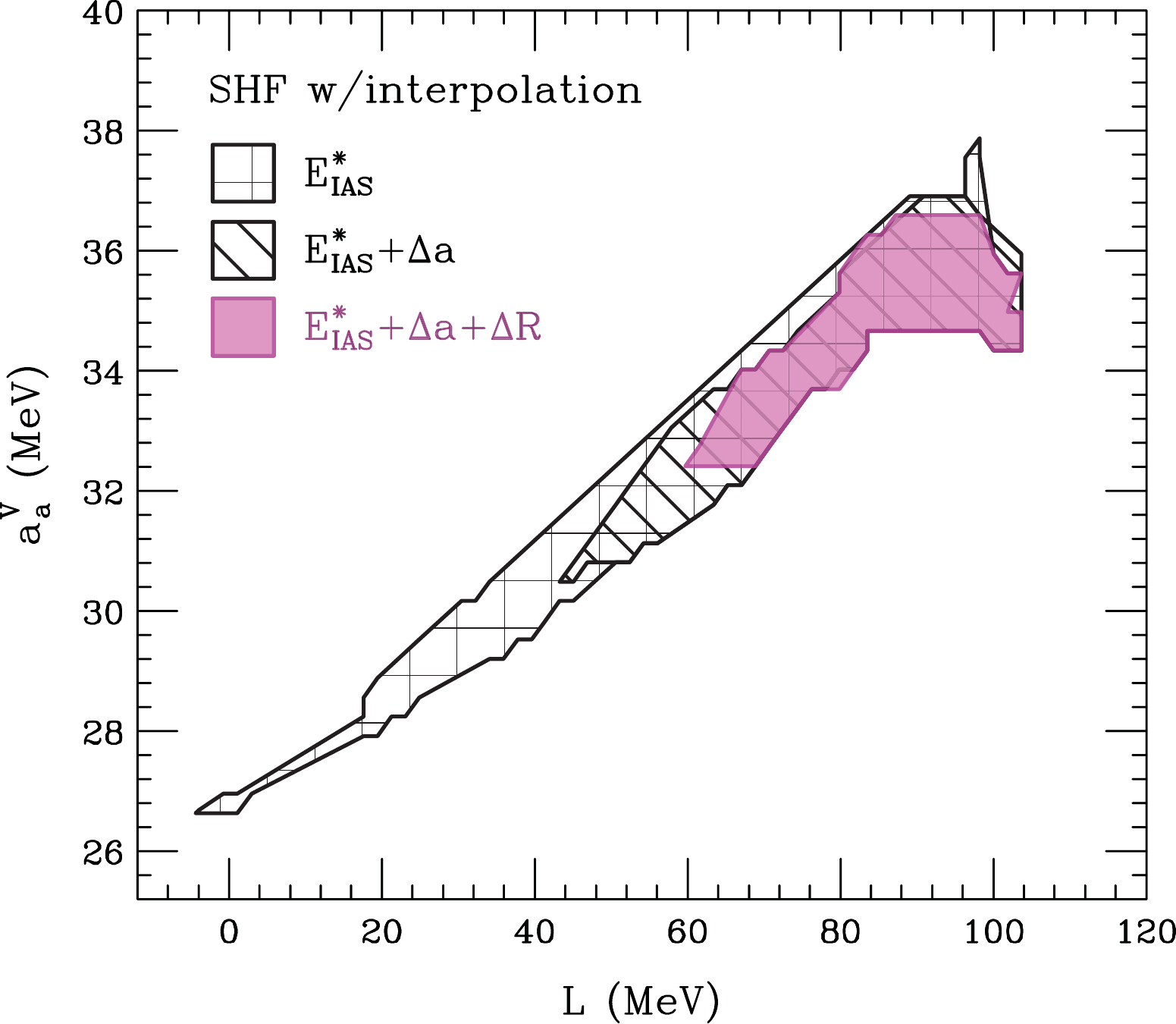}}
\caption{Contours of constant probability density, encircling 68\% of net probability, in the plane of symmetry-energy parameters at normal density, $L$ and $a_a^V$, when more and data is used to improve the inference.  The outmost contour, with interior crossed diagonally, represents the probability obtained when incorporating results from II on the mass-dependence of symmetry-energy coefficients, following from systematics of excitation energies to ground-state IAS, $E^*_\text{IAS}$.  The more inner contour, with interior cross-hatched, represents the probability arrived at after incorporating the result on the average difference of slopes for the isovector and isoscalar potential surfaces, $\Delta a$, cf.~Table~\ref{tab:BestFit}.  The~most inner contour, with shaded interior, represents the probability arrived at after incorporating the results on the differences in the radii of isovector and isoscalar potentials for the four nuclei, $\Delta R$, cf.~Table~\ref{tab:BestFit} again.
}
\label{fig:aavl}
\end{figure}

In Fig.~\ref{fig:aavl}, we show the evolution of the density of probability in the $(L,a_a^V)$ plane~\cite{tsang_constraints_2012} as more and more information from data analyses is accounted for.  Compared to II, we include more Skyrme parametrizations, but now we suppress, in the prior density, those with incompressibilities outside of the realistic region of $200 < K < 300 \, \text{MeV}$, with a~stronger suppression factor the farther they depart from that region.  Even though we interpolate between the interactions and normalize the prior probability density to a constant value within that portion of the $(L,a_a^V)$ plane where pertinent Skyrme parametrizations are available, the~boundaries of the covered region are expectedly noisy and connections between observable quantities and symmetry-energy parameters exhibit graininess tied to the finite number of the starting Skyrme parametrizations.  In spite of those limitations, it is very obvious in Fig.~\ref{fig:aavl} that the reproduction of the mass dependence of symmetry-energy coefficients in~II, $a_a^V(A)$, or equivalently the systematics of $E_\text{IAS}^*$, imposes a narrow positive correlation within the plane of $(L,a_a^V)$.  Within that correlation region, the expectation of reproducing the difference in the slopes $\Delta a$ of isovector and isoscalar surfaces, deduced by fitting the elastic and QE (p,n) cross sections, shrinks the region compatible with data to higher values of symmetry-energy parameters, with $L$ above $\sim 50 \, \text{MeV}$.  Inclusion of consistency
with the differences $\Delta R$ in the isovector and isoscalar radii for the four nuclei, from fitting the reaction data, shrinks the region of likely parameters of the symmetry energy farther out into high values, with $L$ above $\sim 70 \, \text{MeV}$.

\begin{figure}
\centerline{\includegraphics[width=.7\linewidth]{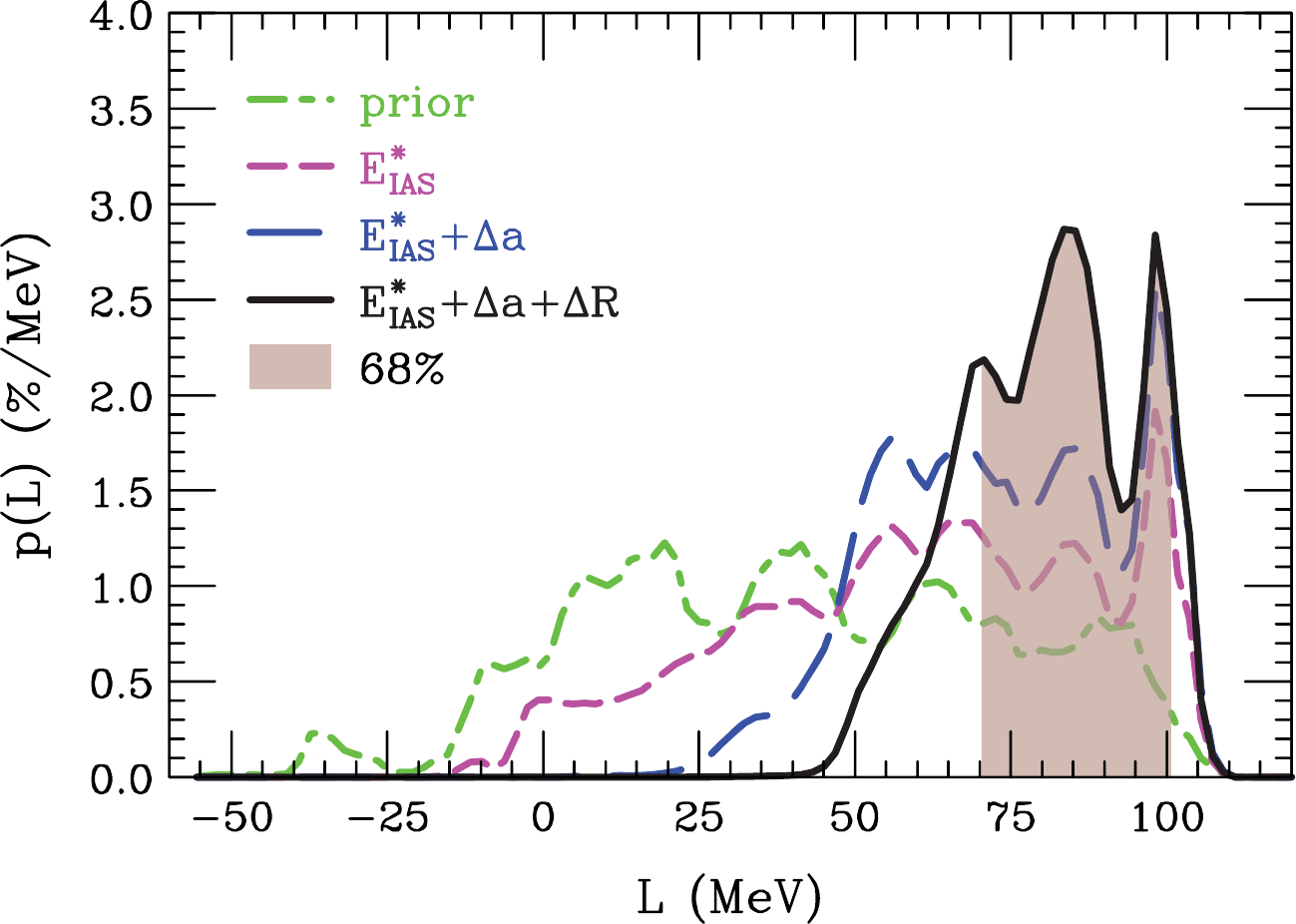}}
\caption{Evolution of the density of probability in the value $L$ of the slope of symmetry energy at $\rho_0$, as different data are accounted for.  The prior density represents the projection of the prior density in the plane of $(L,a_a^V)$, that is normalized to a constant value in the regions of the plane for which pertinent Skyrme parametrizations or their linear combinations can be found.  Oscillations in the prior represent a discontinuous behavior of the boundaries of the covered region in the $(L,a_a^V)$ plane.  The shaded portion
of the final density, with all considered data included, represents the most narrow region in $L$ that contains 68\% of the probability.
}
\label{fig:pl}
\end{figure}

\begin{figure}
\centerline{\includegraphics[width=.7\linewidth]{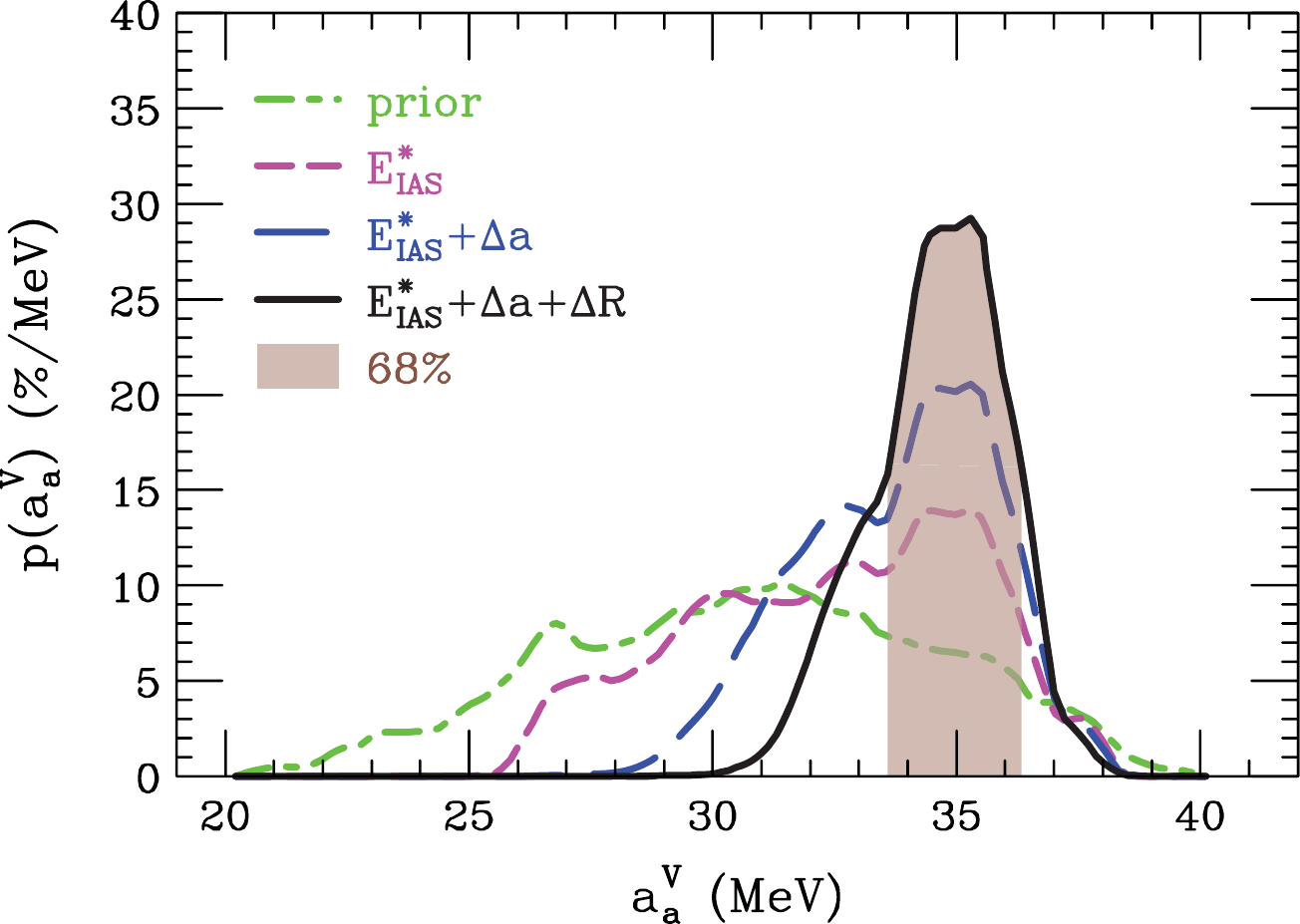}}
\caption{Evolution of the density of probability in the value of the symmetry energy at $\rho_0$, $a_a^V \equiv S(\rho_0)$,  as different data are accounted for.  The prior density represents the projection of the prior density in the plane of $(L,a_a^V)$, that is normalized to a constant value in the regions of the plane for which pertinent Skyrme parametrizations or their linear combinations can be found.  The~shaded portion
of the final density, with all considered data included, represents the most narrow region in $a_a^V$ that contains 68\% of the probability.
}
\label{fig:pv}
\end{figure}

In Figs.~\ref{fig:pl} and \ref{fig:pv}, we show the evolution of the probability density, with addition of data, when projected onto the $L$ and $a_a^V$ axes.  We maintain the same normalization of the prior, to a constant value in the region of the $(L,a_a^V)$ plane that is covered and a~smooth, at the level of our  discretization in the plane, transition to zero for the regions that are not covered.  In Fig.~\ref{fig:pl}, in particular, oscillations are seen tied to noisy boundaries of the covered region.  The oscillations persist after data are incorporated, due to graininess in the connection between symmetry-energy parameters and observables.  As such they should be considered as artifacts of methodology, rather than of significance.

Within Bayesian inference, it may be seen in Figs.~\ref{fig:pl} and~\ref{fig:pv} that even accounting for the systematics of nuclear symmetry-energy coefficients pushes alone the probability towards higher values of $L$ and $a_a^V$.  In~$L$, in fact even a strong push is seen against the high-end boundary of the prior.  That boundary is a reflection of the limitation of the Skyrme parametrizations - relativistic mean field approaches tend to yield higher values of~$L$ and may eventually get invoked to improve inferences on the probability from the high-$L$ side, in the context of the specific set of observables.  By the time the differences in the radii $\Delta R$ are accounted for in the inference, though, that push against the upper $L$-boundary subsides, though, see Fig.~\ref{fig:pl}.  Looking for the most narrow interval along either the $L$- or the $a_a^V$-axis, that contains 68\% of the probability, we find that with this probability the slope and the value are within the limits of $70 < L < 101 \, \text{MeV}$ and $33.5 < a_a^V < 36.4 \, \text{MeV}$, respectively.  The specific value of 68\% is obviously taken because this is the net probability within one error from the central value for a probability density in Gaussian form.

\begin{figure}
\centerline{\includegraphics[width=.7\linewidth]{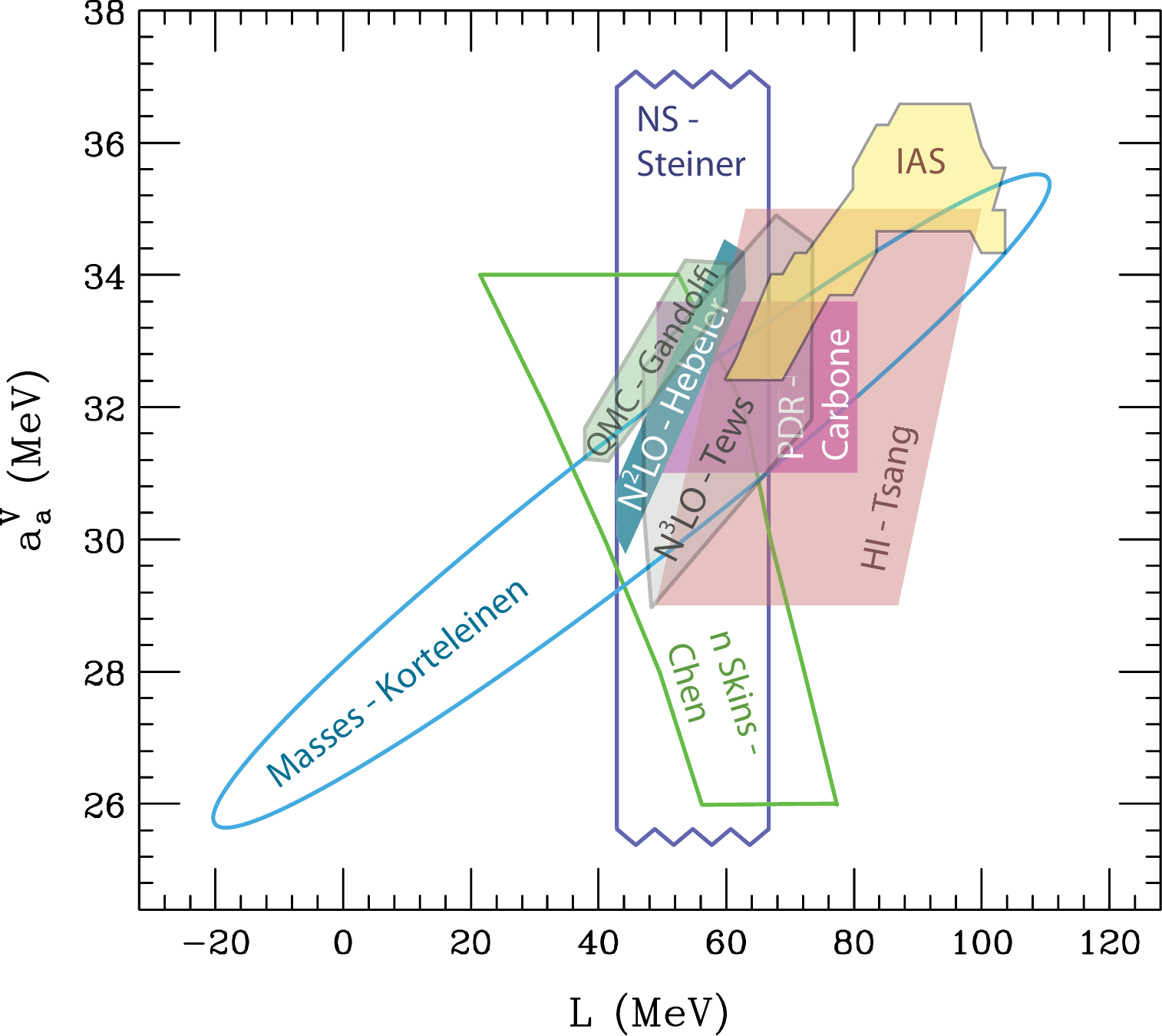}}
\caption{Constraints, from different sources, on the symmetry-energy parameters at $\rho_0$, $a_a^V \equiv S(\rho_0)$ and slope $L$, after \cite{tsang_constraints_2012,lattimer_constraining_2012}.  Included are predictions from neutron-matter calculations by Gandolfi~\etal~\cite{gandolfi_maximum_2011}, within QMC, and within chiral effective field theory, in N$^\text{2}$LO order by Heberle \etal~\cite{hebeler_constraints_2010} and in N$^\text{3}$LO order by Tews \etal~\cite{tews_neutron_2013}.  From Fig.~\ref{fig:aavl}, we further reproduce here our own $E^*_\text{IAS}$ + $\Delta R$ constraints, marked IAS, as combining conclusions from the excitation energies and cross sections to the ground-state IAS.  Other illustrated constraints, deduced from observables, include those deduced from neutron skins by Chen \etal~\cite{chen_density_2010}, from nuclear masses by Korteleinen \etal~\cite{kortelainen_nuclear_2010}, pygmy dipole resonance (PDR) by Carbone~\etal~\cite{carbone_constraints_2010}, heavy-ion collisions~(HI) by Tsang~\etal\ and quoted in~\cite{horowitz_way_2014}, and from neutron-star (NS) observations by Steiner \etal~\cite{0004-637X-722-1-33}.
}
\label{fig:aavl18}
\end{figure}

Next, in Fig.~\ref{fig:aavl18} we compare our constraints on $(L, a_a^V)$ to some of those in the literature, with reasonable realism, specifically from nuclear-matter calculations within quantum Monte-Carlo (QMC) \cite{gandolfi_maximum_2011} and chiral effective field theory (CEFT) within N$^\text{2}$LO order~\cite{hebeler_constraints_2010} and N$^\text{3}$LO~\etal~\cite{tews_neutron_2013}, derived from neutron-star observations~\cite{0004-637X-722-1-33}, ground-state masses~\cite{kortelainen_nuclear_2010} and from various reaction \mbox{observables~\cite{chen_density_2010,carbone_constraints_2010,horowitz_way_2014}.} Relative to other inferences in the literature, ours occupy the high-value corner in both $L$ and $a_a^V$.  In microscopic theory, the parameters of the symmetry energy are quite sensitive to the strength of three-nucleon (3N) interactions (primarily $c_3$ constant), implanted on top of the two-nucleon (2N) interactions, with both $a_a^V$ and $L$ increasing as the 3N strength is increased.  In the case of QMC and CEFT N$^\text{2}$LO, the diagonally slanted regions in Fig.~\ref{fig:aavl18} explicitly illustrate the evolution of the parameters as the strength of the 3N interactions is increased.  The~results of microscopic calculations, though, depend also on chosen strategies, with greater corresponding uncertainty the larger the density and the larger the order of the calculations.  In Fig.~\ref{fig:aavl18} it appears that the best compromise between different inferences in the literature is around  $a_a^V \sim 33 \, \text{MeV}$ and $L \sim 60 \, \text{MeV}$, with significant impact of the 3N forces on the parameters then.  Our results alone, however, favor though even higher values of both parameters.

\begin{figure}
\centerline{\includegraphics[width=.7\linewidth]{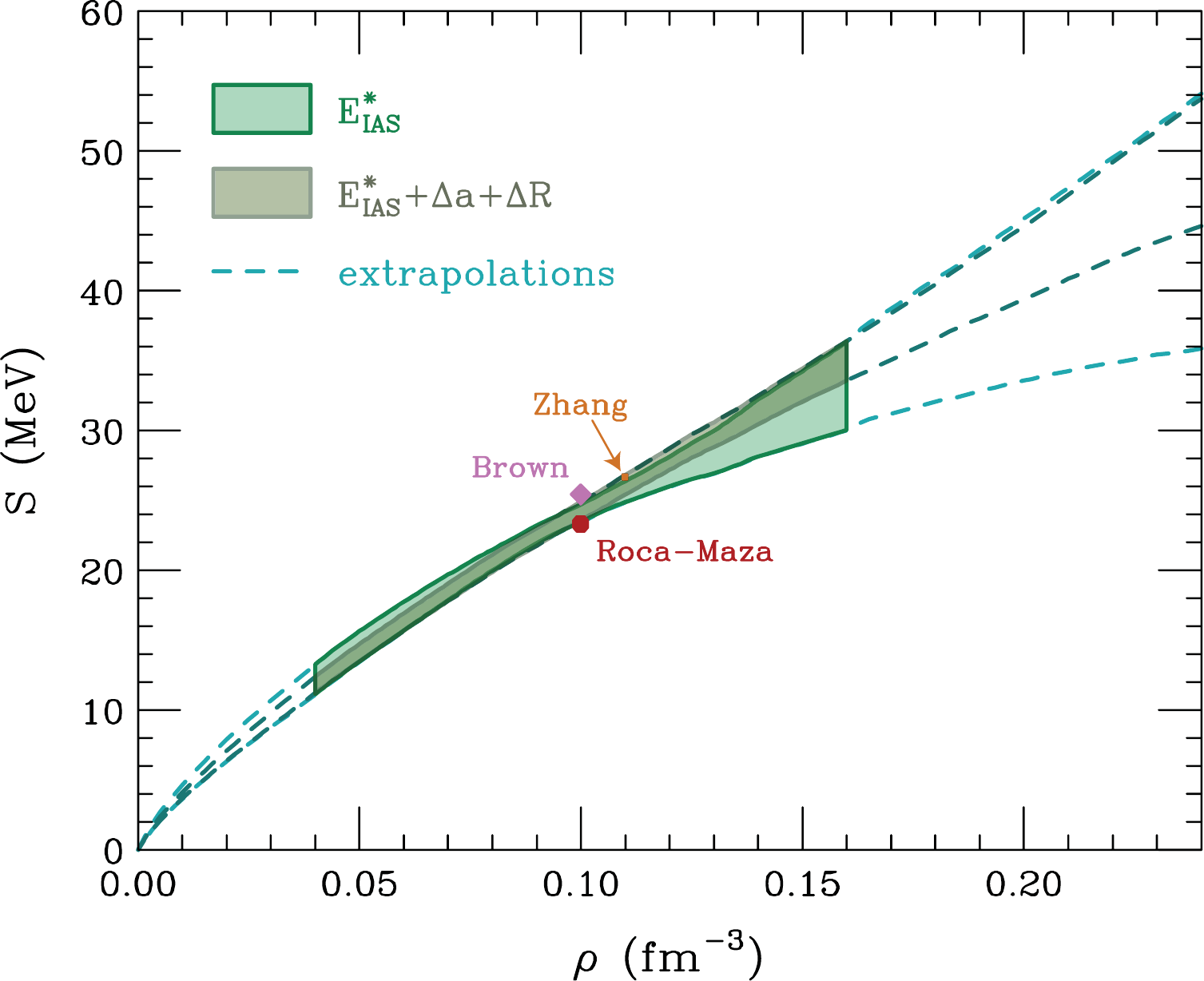}}
\caption{Symmetry energy in uniform matter as a function of density.  The shaded regions represent the most narrow ranges of the symmetry energy at a specific density $\rho$, $0.04 < \rho < 0.16 \, \text{fm}^\text{-3}$, that contain 68\% of probability in the Bayesian inference when SHF calculations are confronted against conclusions from data.  The wider more lightly shaded region, generally encompassing the more narrow darker region, results from data on $E_\text{IAS}^*$ leading to the mass-dependent symmetry-coefficients~$a_a(A)$.  The~more narrow and more darkly shaded region results when the $E_\text{IAS}^*$ systematics is combined with conclusions on difference in geometry for isovector and isoscalar potentials, quantified with $\Delta a$ and $\Delta R$. The~dashed lines represent extrapolations of the described regions to lower and higher densities.  The~three symbols, diamond, circle, and square, represent the values of symmetry energy at moderately subnormal densities deduced, respectively, by Brown~\cite{brown_constraints_2013}, Roca-Maza \etal~\cite{RocaMaza:2012mh}, and by Zhang~\etal~\cite{zhangChen_constraining_2013}.  The~vertical sizes of those symbols represent the claimed errors on the deduced values of the symmetry energy.
}
\label{fig:symecvs}
\end{figure}

Proceeding in a similar manner as in the case of $a_a^V$ and $L$, we next generate limits on the symmetry energy at individual densities, i.e.\ $S(\rho)$, seeking the most narrow intervals that contain 68\% of the probability at specific $\rho$ in Bayesian inference.  We choose to normalize the prior in the same way as before, i.e.\ to a uniform density in the $(L,a_a^V)$ plane, even though the specific analysis does not progress through the density in that plane - in practice this amounts to a specific weighting of the individual Skyrme interactions so they end up giving uniform density of probability in $(L,a_a^V)$.  The~constraints on $S(\rho)$, arrived at when confronting the Skyrme results against conclusions drawn from $E_\text{IAS}^*$, i.e.~on~$a_a(A)$, and when combining the latter conclusions with those from analysing differential cross sections, are shown in Fig.~\ref{fig:symecvs} as different shaded regions for the density interval $0.04 < \rho < 0.16 \, \text{fm}^\text{-3}$.  Only within that density interval the symmetry energy gets in practice tested in a nucleus, cf.~I and~II.  The~wider shaded region in the figure is arrived for fewer data incorporated and the narrower for more.  Notably, this is by no means a rule in the Bayesian inference: a confidence region may widen when consecutive data appear to contradict each other.  The dashed lines show extrapolations of the results to lower and higher densities.  Our credibility regions in Fig.~\ref{fig:symecvs} are bracketed there by values of symmetry energy inferred for intermediate densities from different data, specifically by Brown~\cite{brown_constraints_2013}, Roca-Maza \etal~\cite{RocaMaza:2012mh}, and by Zhang~\etal~\cite{zhangChen_constraining_2013}.  The astounding finding in employing the Bayesian inference, as compared to II, is in the arrival at a generally more narrow credibility region on $S(\rho)$, than in~II, when using just the excitation energies to ground state IAS to narrow the credibility.  The sharp rise in the symmetry energy with density is favored within the $E_\text{IAS}^*$-analysis as explaining a strong variation of the symmetry coefficients with nuclear mass, cf.~II.  Addition of information from analyzing the differential cross sections in the current paper only mildly improves the inference at intermediate to low densities.  In the weakly subnormal density region, the last addition shifts the probability to the stiffest symmetry energies.  It should be stressed, for the perspective, that the prior distribution is fairly broad at all individual densities within the range $0.04 < \rho < 0.16 \, \text{fm}^\text{-3}$, just as at $\rho_0$ in Fig.~\ref{fig:pv}.

\begin{figure}
\centerline{\includegraphics[width=.7\linewidth]{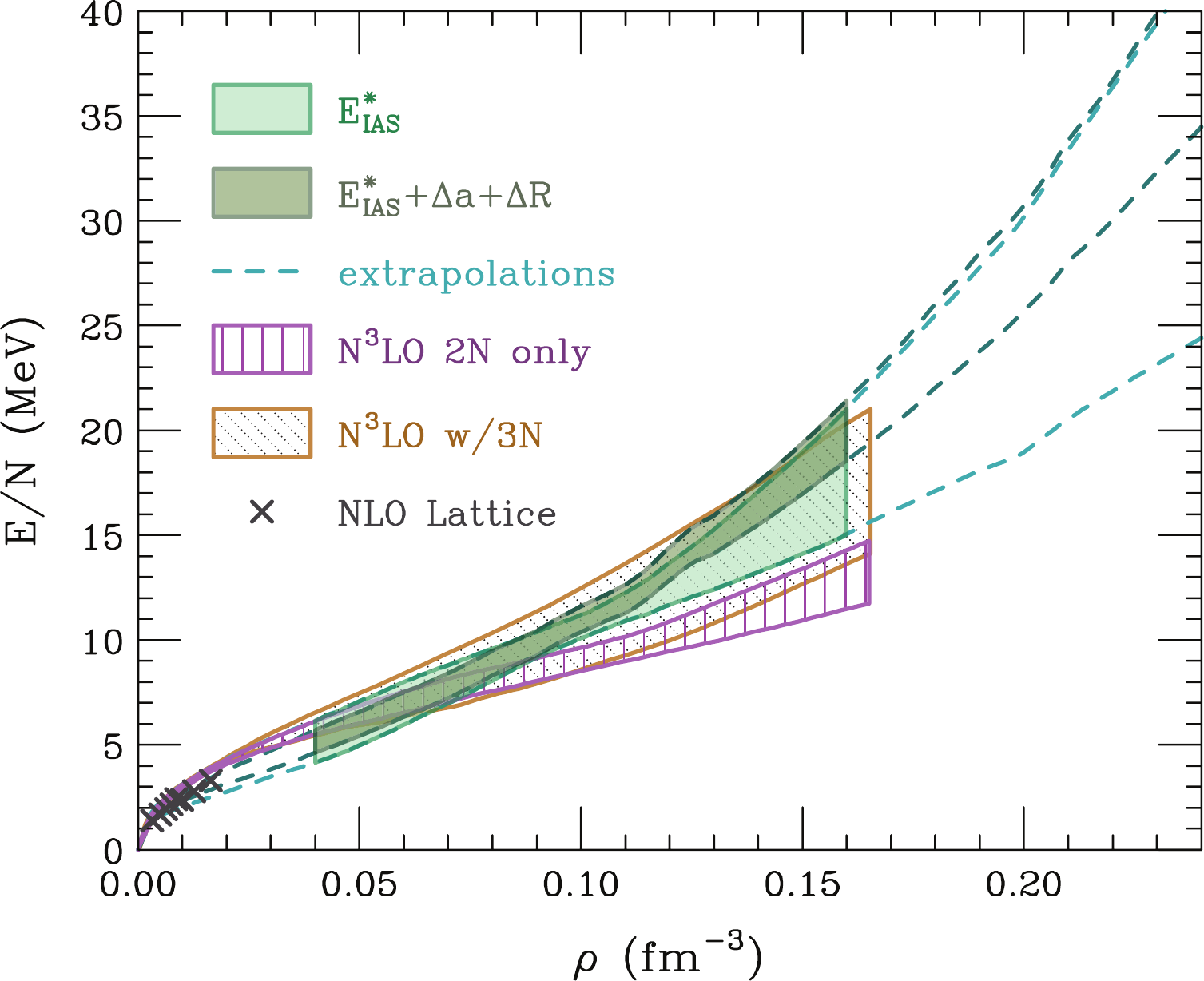}}
\caption{Energy per neutron in uniform neutron matter as a function of density.  The shaded regions represent the most narrow ranges of the energy at a specific density $\rho$, $0.04 < \rho < 0.16 \, \text{fm}^\text{-3}$, that contain 68\% of probability in the Bayesian inference when SHF calculations are confronted against conclusions from data.  The wider and more lightly shaded region, mostly encompassing the more narrow darker region, stems from the data on $E_\text{IAS}^*$ leading to the mass-dependent symmetry-coefficients~$a_a(A)$.  The~more narrow and more darkly shaded region results when the $E_\text{IAS}^*$ systematics is combined with the conclusions on differences in geometry for isovector and isoscalar potentials, quantified with $\Delta a$ and $\Delta R$. The~dashed lines represent extrapolations of the described regions to the lower and higher densities.
Further represented in the figure are predictions from the chiral effective field theory, specifically from the NLO lattice calculations by Epelbaum~\etal~\cite{epelbaum_ground-state_2009} - as crosses, and from the N$^\text{3}$LO calculations by Tews~\etal~\cite{tews_neutron_2013} with 2N interactions only - as~vertically hatched region, and with 3N interactions - as a region hatched diagonally with dots.
}
\label{fig:eaneu}
\end{figure}

Progressing as with $S(\rho)$, we proceed further to arrive at constraints on energy per neutron $\frac{E}{N}(\rho)$ in pure neutron matter, continuing to use the SHF calculations in connecting the theoretical inputs to observables.  As the observables exploited here primarily constraint the symmetry energy portion of $E/N$, we must rely on the expectation that the Skyrme parameterizations on the average reasonably describe the energy per nucleon in symmetric matter and they provide a reasonable spread of values for the quartic term in the expansion in asymmetry for the energy, that leads to the symmetry energy.   I.e.\ we need to rely now more heavily on the sensibility of the prior probability than in the case of $S(\rho)$.  The~results for $\frac{E}{N}(\rho)$, when applying Bayesian inference, are shown in Fig.~\ref{fig:eaneu}.  The lighter and darker shaded regions, in density interval of $0.04 < \rho < 0.16 \, \text{fm}^\text{-3}$, represent 68\% credibility intervals when, respectively, accounting for the $E^*_\text{IAS}$ systematics alone and that systematics in combination with the differences in geometry for the isoscalar and isovector potentials.  The dashed lines in the figure show extrapolations to the lower and higher densities.  Just as with $S(\rho)$, the large isovector skins favor a rapid rise of $\frac{E}{N}(\rho)$ with $\rho$.  Besides our inferences, represented in Fig.~\ref{fig:eaneu} are the CEFT predictions, specifically from NLO lattice calculations Epelbaum~\etal~\cite{epelbaum_ground-state_2009}  - as crosses, and from N$^\text{3}$LO calculations by Tews~\etal~\cite{tews_neutron_2013} with 2N interactions only - as~vertically hatched region, and with 3N interactions - as a region hatched diagonally with dots. (The four-nucleon interactions are also included in the latter calculations, but yield small contributions even at the highest considered densities.)  At~lower densities in the figure, the different CEFT agree between each other and with our inferences.  As density increases, however, the predictions that only rely on 2N interactions rise, though, too slowly compared to the inferences from data.  Those that include 3N interactions overlap with the inferences, but when the isovector skins are accounted for, only the high-energy end of the latter predictions is consistent with the inferences.  In the Tews~\etal\ calculations~\cite{tews_neutron_2013}, that end is built up by the calculations that rely on the Entem-Machleit 2N interactions~\cite{entem_accurate_2003}  within the full interaction set.

\subsection{Neutron Skins}

\begin{figure}
\centerline{\includegraphics[width=.7\linewidth]{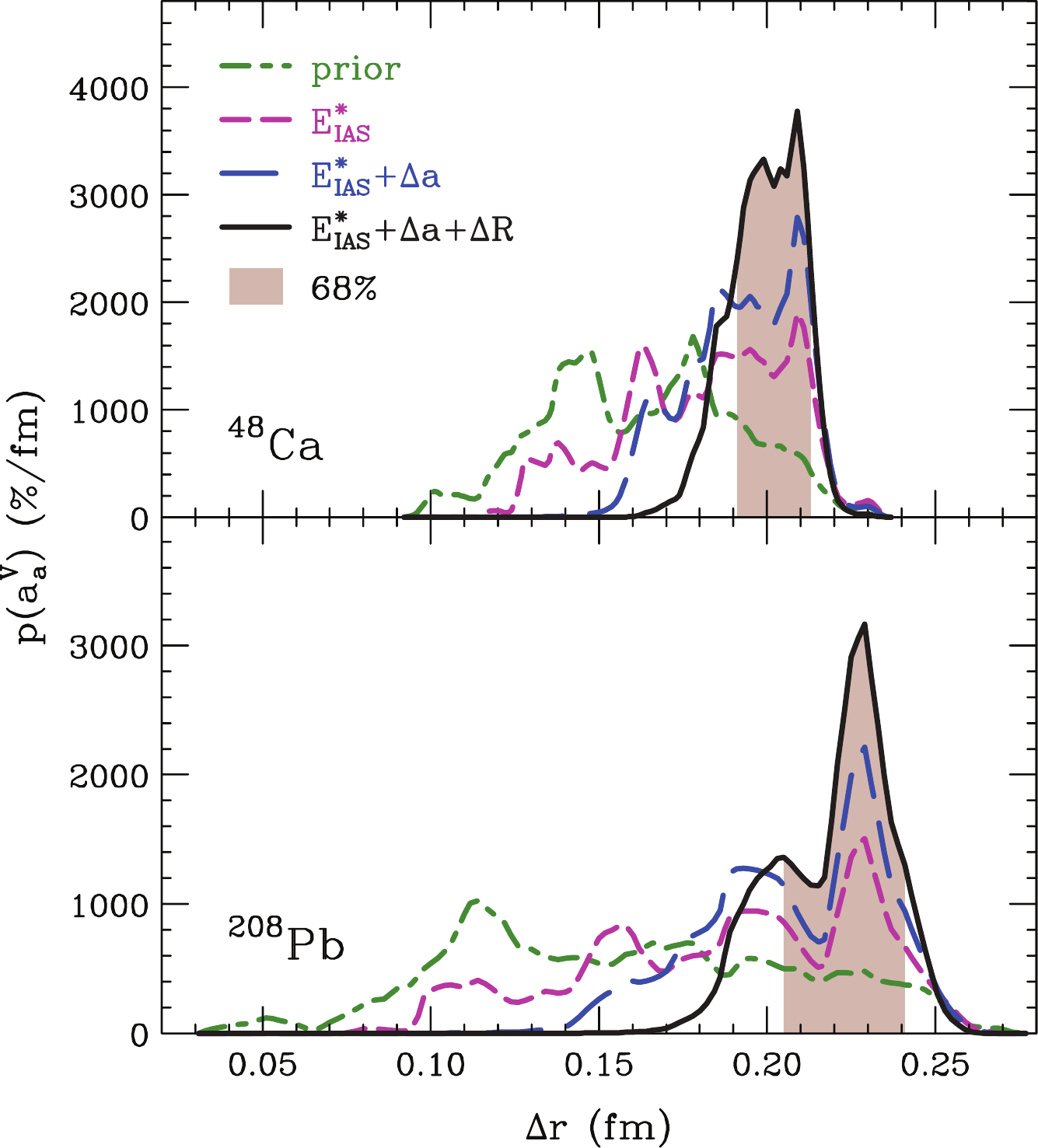}}
\caption{Evolution of the density of probability in the value of neutron skin, $\Delta r = \langle r^2 \rangle_n^{1/2} - \langle r^2 \rangle_p^{1/2}$, for $^\text{48}$Ca (top panel) and $^\text{208}$Pb (bottom), as different data are accounted for.  Oscillations in the prior density are due to the fact that the prior is normalized to uniform in the plane of $(L,a_a^V)$ over a region with uneven boundaries. Oscillations in the posterior density are also tied to a finite number of original Skyrme parametrizations tying $\Delta r$ to specific observables.  The shaded portion
of the final density, with all considered data included, represents the most narrow region in $\Delta r$ that contains 68\% of the probability.
}
\label{fig:PRNP}
\end{figure}

The combination of $L$ and $a_a^V$, that we arrive at in the present work, using the isovector skins as observables, clearly represent stiffer symmetry energies than those arrived at using the neutron skins, see Fig.~\ref{fig:aavl18}, and Ref.~\cite{zhangChen_constraining_2013} or II.  With the position of isovector surface, relative to isoscalar, testing sensitively the stiffness of the symmetry energy, as illustrated in Fig.~\ref{fig:rho_all}, the QE (p,n) charge exchange reactions principally test the stiffness better than the elastic reactions, cf.~Fig.~\ref{fig:ppn48Ca35_fvr}.  The latter reactions were used in the past to determine the neutron skins.  Even in other analyses of data, it has been common to transcribe any conclusions onto the skin values, most often for $^\text{208}$Pb.  Another neutron skin of interest in the literature recently has been that of $^\text{48}$Ca.  Earlier in this work, we estimated that the isovector skins differ from neutron skins by a factor of few for these specific nuclei.  Clearly of interest can be what the current conclusions yield in detail.  Evolution of the probability in the neutron skin values, $\Delta r = \langle r^2 \rangle_n^{1/2} - \langle r^2 \rangle_p^{1/2}$, as different data are incorporated, is~illustrated in Fig.~\ref{fig:PRNP}.  In the top panel the evolution is shown for the lighter of the two nuclei and in the bottom -- for the heavier.  Looking for the most narrow intervals that contain 68\% of probability, we arrive at the confidence limits of $0.191 < \Delta r < 0.213 \, \text{fm}$ and $0.205 < \Delta r < 0.241 \, \text{fm}$, for $^\text{48}$Ca and~$^\text{208}$Pb, respectively.  These intervals lie above those arrived at in II for the two nuclei when summing up various analyses of data directly pertaining to the skins, primarily from elastic scattering.  (The analysis of a later pion photoproduction experiment on $^\text{208}$Pb~\cite{crystal_ball_at_mami_and_a2_collaboration_neutron_2014} though, not included in II, gave as well a skin compatible with those in~II.)  Regarding theoretical predictions, the CEFT one \cite{hagen_neutron_2016} of $0.13 < \Delta r < 0.15 \, \text{fm}$ for $^\text{48}$Ca is particularly low compared to the range inferred here and in the context of the density of probability in Fig.~\ref{fig:PRNP}, though not inconsistent with the analyses summed up in II.  However then the symmetry energies for the interactions touted in \cite{hagen_neutron_2016} are particularly soft, with $37.8 < L < 47.7 \, \text{MeV}$ and $ 25.2 < a_a^V < 30.4 \, \text{MeV}$.  The specific interaction promoted there, NNLO$_\text{sat}$, is in fact represented in the lower halves of these intervals and not realistic even in the context of the literature claims on $\Delta r$, summarized in II.

\subsection{Droplet-Model Context}

The size of isovector skins supported by data may seem startlingly large.  Irrespectively of amplification factors relative to neutron skins and results from structure, such as in Fig.~\ref{fig:rho_all} where the skins evidence gets marred by shell effects, it can be important to gain an additional perspective as to whether the large skins make sense or not.  Such a perspective may be provided by macroscopic models such as those used to describe average features of nuclear masses~\cite{myers_nuclear_1974,moller_nuclear_1995,Danielewicz:2003dd}.

Within a macroscopic model~\cite{Danielewicz:2003dd}, portion of neutron-proton imbalance is pushed out to the nuclear surface, reducing the energetic penalty for the imbalance in the interior.  The ratio of the imbalance in the surface to the interior scales in proportion to surface and volume capacitances for asymmetry:
\beq
\frac{(N-Z)_S}{(N-Z)_V} = \frac{C_S}{C_V} = \frac{A^{2/3}/a_a^S}{A/a_a^V} = \frac{a_a^V}{a_a^S \, A^{1/3}} \, .
\eeq
On the other hand, we can write for the ratio of the surface-to-volume asymmetries:
\beq
\frac{(N-Z)_S}{(N-Z)_V} \simeq \frac{4 \pi \, r_0^2 \, A^{2/3} \, \Delta R}{(4\pi/3) \, r_0^3 \, A} = \frac{3 \Delta R}{r_0 \, A^{1/3}} \, .
\eeq
In the above, we assumed approximately the same isovector density in the interior and in the surface.  Combining the two results we arrive at (see also I)
\beq
\frac{a_a^V}{a_a^S} \simeq \frac{3 \Delta R}{r_0} \simeq 2.3 \,.
\eeq
In the last step, we used the global fit value of $\Delta R = 0.88 \, \text{fm}$ from Table \ref{tab:BestFit1} and $r_0=1.14 \, \text{fm}$.  The ratio of volume-to-surface symmetry coefficients of 2.3, equivalent to the surface skin $\Delta R \sim 0.88 \, \text{fm}$, is fairly mundane as far as the droplet-model considerations are concerned~\cite{myers_nuclear_1974,moller_nuclear_1995,Danielewicz:2003dd}.

A more accurate mapping of the current results onto the $(a_a^V,a_a^S)$ pair would require another Bayesian inference, beyond the scope of the present paper.  We should mention that when the ratio $a_a^V/a_a^S$ is fitted to masses or excitation energies to IAS, an inflated value may emerge compared to $A \rightarrow \infty$ limit, cf.~II.
According to Fig.~\ref{fig:drx}, the geometric vector skins seem to approach the macroscopic limit much faster then the surface details in masses.

\section{Conclusions}

In this work, we simultaneously analyzed differential cross sections for elastic (p,p) and (n,n) reactions, and quasielastic (p,n) reactions to IAS, on four targets, $^{48}$Ca, $^{90}$Zr, $^{120}$Sn and $^{208}$Pb, within the energy range of (10--50)~MeV, following the concepts of isoscalar and isovector potentials combined into a Lane potential.  The goal was to detect and quantify a~possible displacement of the isovector and isoscalar surfaces suggested by the macroscopic considerations and by structure calculations.  For this purpose, we minimally modified the popular Koning-Delaroche potential to allow for controlled changes in the geometry of the isovector and isoscalar potentials.  We demonstrated that the geometry of isovector potentials strongly impacts differential charge-exchange cross sections and only weakly elastic. Accordingly, we organized the fit procedure so that the elastic cross sections primarily governed the adjustments in the isoscalar potentials and the charge-exchange cross sections -- the adjustments in the isovector potentials relative to isoscalar.  Fits to the data gave rise to large differences $\sim$$0.9 \, \text{fm}$ in the radii of isovector and isoscalar potentials, or isovector skins, changing little from a nucleus to a nucleus.  In addition, the isovector surfaces were found to be slightly steeper, by $\sim$$0.1 \, \text{fm}$ in diffusivity, compared to isoscalar.  The particular aspect of the charge-exchange cross sections that is sensitive to the isovector skins are the oscillations as a function of angle, strengthening and changing position as skin size increases.

In the impulse approximation and in the folding models, the isovector and isoscalar potentials are tied to the isovector and isoscalar densities, respectively.  In the structure calculations and in Thomas-Fermi considerations, average differences between the latter densities are tightly tied to the density dependence of the symmetry energy.  Expecting the differences in the geometry of the potentials to reflect the differences in the geometry of underlying densities, we attempted to use the differences from the fits to learn on the symmetry energy.  Large isovector skins, such as found in analyzing cross sections, roughly independent of a~nucleus, with a steeper isovector surface than isoscalar are produced in Skyrme-Hartree-Fock calculations using relatively stiff symmetry energies.  Relying on the current data analysis and on results of our prior work examining data on excitation energies to isobaric analog states of ground states, and employing Bayesian inference, we arrived at 68\% credibility limits for the parameters of symmetry energy at normal density, of
$70 < L < 101 \, \text{MeV}$ and $33.5 < a_a^V < 36.4 \, \text{MeV}$.  We also arrived at credibility regions on symmetry energy and energy of neutron matter as functions of density.  From the results on neutron matter in chiral effective field theory only those most stiff overlap with our constraint region.  When transcribing our isovector skin results into neutron skins, within structure calculations, we arrived at large values, in particular $0.205 < \Delta r < 0.241 \, \text{fm}$ for $^{208}$Pb.

The greatest weakness of the analysis here is the attribution of the differences in the geometry of optical potentials, needed to explain differential cross sections, to the differences in the densities.  Weak dependence of the inferred neutron skin on the nucleus seems to support this attribution.  We should add that the Coulomb energy for a proton by the surface of the target strongly changes between $^{48}$Ca and $^{208}$Pb, so effectively we arrive at similar results for different energy brackets of a proton by the target nucleus.

Further work on this topic could be improved in different directions.  Thus, the analysis critically depends on availability of quasielastic (p,n) data with a good resolution of oscillations in the differential cross sections.  Data at high incident energies are usually favored investigations of densities.  Fitting of potential parameters here was done following frequentist methodology and it is Bayesian methodology that is usually particularly effective when data of different type, such as from (p,p), (n,n) and (p,n) processes are combined.  We, so far, did not exploit analyzing power in the analysis.  In the paper we indicated that the isovector potentials multiplying the third and transverse isospin components should be different due to the Coulomb polarization, but we did not exploited that in the data analysis.  On the structure side, we were short in the interactions with very stiff symmetry energies, such as common in relativistic mean field calculations.

\acknowledgments

Strong impetus for this work was provided by the talk by Dao Tien Khoa given in the context of the 2013 International Collaborations in Nuclear Theory Program on Symmetry Energy.  In~the early stages of this work, the authors benefited from the expertise of Luke Titus.  As work progressed, the authors benefited from discussions with many colleagues including Remco Zegers, Brent Barker, Ron Johnson, Achim Richter, Sam Austin, Naftali Auerbach and Bill Lynch.  Alex Brown provided the authors with a family of Skyrme interaction parameters \cite{brown_constraints_2013}.  Progress was further boosted through attendance of workshops co-sponsored by CUSTIPEN in 2015, on~reactions and on asymmetric nuclear matter.
This work was supported by the U.S.\ National Science Foundation under Grants PHY-1068571 and PHY-1403906 and by the University Grants Commission of India under Indo-US 21$^\text{st}$ Century Knowledge Initiative.

\newpage

\bibliography{rsv16,ss11,skyrme,skin}

\bibliographystyle{my}  

\end{document}